\documentclass[amsmath,12pt,amssymb,preprint,nofootinbib]{revtex4-2}
\usepackage[justification=raggedright,singlelinecheck=false]{caption}
\usepackage{amsfonts} %\usepackage{citesort}
\usepackage{graphicx} % Include figure files
\usepackage{epsfig}
\usepackage{multirow}
\usepackage{bm}% bold math
\usepackage{color}
\usepackage{amssymb}
\usepackage{dsfont}
\usepackage{hyperref}
\usepackage{cleveref}
\usepackage{caption}
\usepackage{subcaption}

\usepackage{placeins}  
\usepackage{multirow}
\usepackage{rotating}
\usepackage{array}
\newcolumntype{V}{!{\vrule width 1.5pt}}


\newcommand{\be}{\begin{eqnarray}}
\newcommand{\ee}{\end{eqnarray}}

\usepackage{array}
\newcolumntype{V}{!{\vrule width 1.5pt}}
\begin{document}
\title{Magnetic Moment of Octet Baryons in Isospin Asymmetric Magnetized Strange Matter}
\author{Akshada Waman}
\email{akshadawaman2003@gmail.com}
\author{Priyanshi}
\email{priyanshig2099@gmail.com}
\author{Suneel Dutt}
\email{dutts@nitj.ac.in}
\author{Harleen Dahiya}
\email{dahiyah@nitj.ac.in}

\author{Arvind Kumar}
\email{kumara@nitj.ac.in}

\affiliation{Department of Physics, Dr. B. R. Ambedkar National
	Institute of Technology, Jalandhar, 144008, India}

	\begin{abstract}
We investigate the magnetic moments of octet baryons in isospin asymmetric strange hadronic matter under strong external magnetic fields within a unified theoretical framework by combining the chiral SU(3) quark mean field (CQMF) model with the chiral constituent quark ($\chi$CQM) model. At finite temperature, the inclusion of Dirac sea (DS) effect leads to magnetic catalysis attributing to the enhancement of scalar condensates with increasing magnetic field strength. As a consequence, the effective masses of the octet baryons exhibit a monotonic increase as a function of magnetic field. The results highlight the crucial role of vacuum polarization effects in determining the electromagnetic properties of baryons in strongly magnetized matter having  relevance in heavy-ion collision and compact stars. 
\end{abstract}

	\maketitle
\newpage	
	% \section{\label{intro}Introduction}
\section{Introduction}

% {from reference}

% The cosmological background fields can reach up to 100-200 mpi.\^2 which puts a huge impact on the matter present in neutron stars and other astrophysical objects, whilst the fields produced in RHIC and LHC are of lower magnitude.
Understanding the internal structure of hadrons and how they modify under extreme conditions remains a fundamental problem in nuclear and particle physics. At low energies, the strong interaction is governed by the non–perturbative regime of Quantum Chromodynamics (QCD), where quark confinement and spontaneous chiral symmetry breaking determine the basic properties of hadrons. These mechanisms are responsible for generating the effective masses of the constituent quarks and for the formation of hadrons. When strongly interacting matter is subjected to extreme conditions such as high temperature and/or finite baryon density ($\rho_B$), as well as strong magnetic fields, the underlying QCD vacuum structure is significantly modified and chiral symmetry is expected to be partially restored \cite{Nambu:1961tp,Gross:1973id,Fukushima2011,Shuryak2004,Thomas1984}. Strong magnetic fields are expected to be generated in non-central relativistic heavy-ion collisions, where the fast moving spectator protons generate intense electromagnetic fields during the early stages of the collision. Model calculations indicate that the magnetic field strength can reach up to $eB \sim 3m_\pi^2$ at the Relativistic Heavy Ion Collider (RHIC) \cite{Skokov:2009qp} and up to $eB \sim 15m_\pi^2$ at the Large Hadron Collider (LHC) \cite{Deng:2012pc,Huang:2015oca,Bzdak:2011yy,Gezhagn:2021oav}. These magnetic fields are comparable to intrinsic QCD energy scales and can significantly influence strongly interacting matter through mechanisms such as Landau quantization and magnetic catalysis. 
%  In the relevant literature, people usually use m2 π or MeV2 as the unit of eB where e is the electron charge magnitude
% and mπ ≈ 140 MeV is the pion mass. In converting to the SI or Gaussian units, it is helpful to note the following
% relation: 1 MeV2 = e·1.6904 ×1014 Gauss (if ¯h = c = 1, otherwise the right-hand side should be multiplied by ¯hc2)
In addition, the high electrical conductivity of the quark-gluon plasma can sustain induced currents, thereby slowing down the decay of the magnetic field and allowing it to influence the evolution of the medium \cite{Tuchin:2014hza,Gusynin:1995nb,Andersen:2014xxa,Tuchin:2012mf,McLerran:2013hla}. Such conditions are also associated with anomalous transport phenomena, including the chiral magnetic effect \cite{Kharzeev:2015znc,Fukushima2008,Hirono:2014oda}. Such intense magnetic fields are not restricted to heavy-ion collisions but are also expected in astrophysical environments such as magnetars, where the surface magnetic fields can reach $10^{14}$--$10^{15}$~G, with even stronger fields in their interiors \cite{Duncan1992,Kouveliotou:1998ze,Turolla:2015mwa,Woods:2004kb}. In these compact stars, matter exists at extremely high densities and may contain strange baryons. Under such conditions, baryon magnetic moments are expected to contribute to the magnetization of dense matter and may influence both the equation of state and the macroscopic properties of neutron stars \cite{Broderick:2000pe,Dexheimer:2011pz,Rabhi:2009ih}. As a result, hadronic properties such as masses, magnetic moments, and electromagnetic form factors undergo substantial modifications, providing important insight into the internal quark dynamics and non-perturbative nature of QCD \cite{Papazoglou1998,Ryu:2010bu,Menezes:2008qt,Singh:2016hiw,Ramalho:2025kii}.
%%%%% edited

% \cite{Fukushima2008,Skokov:2009qp,Deng:2012pc}. Under such conditions, hadronic properties such as masses, magnetic moments, and electromagnetic form factors can undergo substantial medium modifications, 
% %%%

Among the various observables used to probe the hadron structure, the magnetic moments of baryons provide important insights. Magnetic moments directly reflect the internal spin and orbital motion of the constituent quarks and therefore provide valuable information about the dynamics governing the baryon wave function. In quark-based descriptions, the total magnetic moment of a baryon arises from several sources, including contributions from valence quarks, polarized quark-antiquark pairs in the sea, and orbital angular momentum associated with chiral fluctuations. Because of their sensitivity to the internal quark dynamics, baryon magnetic moments serve as a valuable probe of non-perturbative QCD and flavor symmetry breaking \cite{Coleman1961,Beg1964}, as well as their possible modification in different environments. For this reason, the magnetic moments of the baryon octet have been extensively investigated in vacuum within a variety of theoretical frameworks. Early studies based on the non-relativistic constituent quark (CQM) model  \cite{ChoudhuryJoshi1976} and the MIT bag model \cite{DeGrand1975} provided a simple yet successful description of baryon magnetic moments in terms of constituent quark degrees of freedom. The subsequent work focused on incorporating dynamical aspects of chiral symmetry breaking, with the Nambu-Jona-Lasinio (NJL) model \cite{Nambu1961} providing an effective description in this direction. Building on these ideas, the chiral constituent quark ($\chi$CQM) model was developed where the effects of Goldstone bosons emission from constituent quarks are included \cite{Ramalho2011,Cheng1995,Sharma2010}. Within this framework, quark-sea polarization and orbital angular momentum contributions arise naturally and play a crucial role in reproducing the experimentally observed magnetic moments of octet baryons. These quark-sea contributions can be interpreted as arising from the quark-antiquark pairs, often referred to as the Dirac sea, generated through chiral fluctuations of constituent quarks. In effective chiral approaches, emission of Goldstone bosons generates polarized sea quarks, which play a significant role in understanding spin-dependent and electromagnetic properties of baryons \cite{Gupta2003,Aguirre:2016vqa}. These contributions become particularly important in describing observables such as magnetic moments and form factors beyond the naive constituent quark picture.

As magnetic moments are fundamental observables, it is important to consider the broader electromagnetic structure of baryons, since magnetic moments provide information about the internal spin and charge distribution of baryons, their behavior under strong magnetic fields offers important insight into the modification of baryonic structure in dense hadronic matter. Various approaches like QCD sum rules \cite{Ioffe:1981kw,Lee1998}, chiral perturbation theory \cite{Bernard:2006gx,Bernard1995}, lattice QCD simulations, and covariant quark models, have been used to investigate the electromagnetic structure of baryons. Such approaches have significantly improved our understanding of the role of relativistic effects and non–perturbative dynamics in determining baryon properties. However, most of these investigations have been carried out under vacuum conditions, where medium-induced effects are absent. In realistic physical systems such as nuclear matter or neutron stars, hadrons exist in a strongly interacting environment where their properties can differ substantially from those observed in free space and electromagnetic properties are expected to undergo notable modifications. Experimental evidence accumulated over the past several decades strongly supports the existence of medium modifications in hadronic structure. One of the most prominent experimental examples of medium modification is the European Muon Collaboration (EMC) effect, which demonstrated that the quark structure of nucleons inside nuclei differs from that of free nucleons \cite{EuropeanMuon:1983wih,Geesaman1995}.
% The EMC experiment, performed at CERN in the early 1980s, studied deep inelastic scattering of high-energy muons off various nuclear targets such as iron and compared the results with those obtained from deuterium. The structure functions extracted from these measurements showed that the quark momentum distributions inside nucleons bound in nuclei differ systematically from those of free nucleons \cite{EuropeanMuon:1983wih}. In particular, the per-nucleon structure function ratio of heavy nuclei to deuterium was found to deviate significantly from unity in the intermediate Bjorken-$x$ region ($0.3 \lesssim x \lesssim 0.7$). This observation provided the first clear evidence that the internal quark structure of nucleons is modified when they are embedded in a nuclear environment. Subsequent experimental and theoretical studies have confirmed that nuclear binding effects, nucleon correlations, and modifications of the quark distributions all contribute to this phenomenon \cite{Geesaman1995}.
Subsequent deep inelastic and polarized electron scattering experiments on light and medium-mass nuclei have provided further evidence that the electromagnetic structure of nucleons is modified inside the nuclear medium \cite{Lu:2001mf,JeffersonLabE93-049:2002asn,Malace:2014uea}. These observations indicate that static electromagnetic observables such as magnetic moments are sensitive to the nuclear environment and motivate theoretical studies of hadron properties beyond vacuum conditions.

A number of theoretical approaches have been developed to describe such in-medium modifications. In hadronic models such as the quark–meson coupling (QMC) model, baryons are treated as composite systems of confined quarks interacting with scalar and vector meson fields \cite{Guichon1988,Saito1994,Ryu2010}. These models successfully reproduce nuclear matter saturation properties and predict density-dependent changes in baryon masses. However, the microscopic response of constituent quarks to strong external magnetic fields is not fully resolved within such frameworks, particularly with regard to Landau quantization and charge-dependent quark dynamics \cite{Fukushima:2008wg,Menezes:2008qt}. Similarly, relativistic mean-field (RMF) models and linear sigma models generally treat baryons as point-like particles, incorporating magnetic-field effects through anomalous magnetic moment terms without explicitly resolving quark-level dynamics \cite{Ali2013,Papazoglou1999,Chakrabarty1997,Broderick:2000pe}. More quark-oriented approaches, such as the NJL and Polyakov-NJL (PNJL) models, have been widely used to investigate strongly interacting matter under external magnetic fields \cite{Fukushima:2008wg,Skokov2011,Klevansky:1992qe,Miransky:2015ava}. These approaches have also been extensively employed to study chiral phase transitions, including effects such as magnetic catalysis and inverse magnetic catalysis \cite{Miransky:2002rp,Shovkovy:2012zn,Bali:2011qj,Pagura2017,Bali2012,Bali:2013esa,Bornyakov2014}. However, a key limitation is that they do not incorporate confinement and treat quarks as independent degrees of freedom, restricting their direct application to important baryonic observables such as magnetic moments \cite{Buballa:2003qv}.

Another important feature of dense matter in neutron stars is its large isospin asymmetry. In neutron-rich matter, the imbalance between up and down quarks leads to isospin-dependent modifications of baryon properties. Within effective descriptions of nuclear matter, the $\delta$ meson plays a key role in generating mass splitting between particles with different isospin projections, thereby enhancing isospin effects in dense matter \cite{Kubis:1997ew,Li:2008gp}. When combined with strong magnetic fields and finite strangeness content, isospin asymmetry can lead to significant modifications in the electromagnetic properties of baryons. These considerations highlight the need for theoretical frameworks that consistently account for quark substructure, chiral symmetry dynamics, and medium-dependent effects. In this context, the CQMF model provides a self-consistent description of quark interactions with scalar and vector meson fields, leading to medium-dependent effective quark masses through spontaneous and explicit symmetry breaking \cite{Papazoglou1999,Wang:2001jw,Singh:2016hiw,Kumari:2020mci}. When combined with the $\chi$CQM model, this framework enables the calculation of baryon magnetic moments while consistently incorporating medium-modified quark masses \cite{Dahiya:2018ahb,Gupta2003}. In the present work, we investigate the magnetic moments of octet baryons in isospin-asymmetric magnetized strange matter within a unified CQMF-$\chi$CQM framework.
% In this approach, the effective quark masses obtained from the CQMF model are used as inputs in the $\chi$CQM, allowing the magnetic moments to be decomposed into valence, sea, and orbital contributions 

 % This hybrid approach provides a consistent quark-level description of the combined effects of density, magnetic field strength, isospin asymmetry, and strangeness on baryon magnetic moments, thereby extending previous studies to environments relevant for relativistic heavy-ion collisions and neutron star physics.

The structure of this paper is as follows. In Sec. ~\ref{Section II}, the theoretical framework employed in the present work is discussed. The methodology consists of two parts: the chiral quark mean field (CQMF) model and the chiral constituent quark model ($\chi$CQM). The CQMF model provides the medium-modified quark masses, which are subsequently used within the $\chi$CQM framework. In Sec. ~\ref{Results}, the medium modifications of densities, effective masses, and magnetic moments of octet baryons are analyzed in the presence of strong magnetic fields, isospin asymmetry, and strangeness fraction. Finally, the main conclusions of the present work are summarized in Sec. ~\ref{Summary}.

\section{Methodology}
\label{Section II}
In this section, we present the theoretical framework used to study the magnetic moments of baryons in isospin-asymmetric strange matter in the presence of magnetic field. In subsection A, the CQMF model is used to determine the in-medium quark masses in a self-consistent manner through quark-meson interactions. In subsection B, the $\chi$CQM model is considered for the calculation of magnetic moment by considering the explicit contribution of valence quarks, sea quark polarization and orbital motion to the baryon magnetic moments. Taken together, these approaches allow us to connect quark-level dynamics with observable baryonic properties.

\subsection*{A. The Chiral SU(3) Quark Mean Field Model}

The CQMF model is an effective approach that provides a description of hadronic matter at finite temperature and baryon density. By incorporating chiral symmetry along with explicit scale symmetry breaking, the model generates medium dependent constituent quark and baryon masses while effectively simulating quark confinement properties of baryons \cite{Wang:2001jw}. The model is particularly suited for exploring strange hadronic systems and the dense interiors of neutron stars, where both isospin asymmetry and strangeness play crucial roles.
% \subsubsection*{1. Lagrangian and Symmetry Structure}
The effective chiral SU(3) Lagrangian is expressed as \cite{Papazoglou1999,Wang:2001jw}
\begin{equation}
\mathcal{L} = \mathcal{L}_{\mathrm{q_0}} + \mathcal{L}_{\mathrm{qm}} + \mathcal{L}_{\mathrm{\Sigma\Sigma}} + \mathcal{L}_{\mathrm{VV}} + \mathcal{L}_{\mathrm{\chi SB}} + \mathcal{L}_{\mathrm{mag}},
\label{Lagrangian}
\end{equation}
where, $\mathcal{L}_{\mathrm{q_0}} = \bar \Psi i\gamma^\mu \partial_\mu \Psi$ is the free quark term, $\mathcal{L}_{\mathrm{qm}}$ describes the quark–meson interaction, $\mathcal{L}_{\mathrm{\Sigma\Sigma}}$ and $\mathcal{L}_{\mathrm{VV}}$ represent the scalar and vector meson self-interaction terms, respectively, $\mathcal{L}_{\mathrm{\chi SB}}$ introduces explicit chiral symmetry breaking, $\mathcal{L}_{\mathrm{mag}}$ includes the effect of external magnetic field. The quark–meson interaction term is written as \cite{Wang:2001jw,Kumari:2020mci} 
\begin{align}
\mathcal{L}_{\mathrm{qm}} &=
g_s \left( \bar{\Psi}_L M \Psi_R + \bar{\Psi}_R M^\dagger \Psi_L \right)
- g_v \left( \bar{\Psi}_L \gamma^\mu l_\mu \Psi_L + \bar{\Psi}_R \gamma^\mu r_\mu \Psi_R \right) \nonumber \\
&= \frac{g_s}{\sqrt{2}} \, \bar{\Psi} 
\left( \sum_{a=0}^{8} \sigma_a \lambda_a 
+ i \sum_{a=0}^{8} \pi_a \lambda_a \gamma^5 \right) \Psi \nonumber \\
&\quad - \frac{g_v}{2\sqrt{2}} \, \bar{\Psi}
\left( \sum_{a=0}^{8} \gamma^\mu v_\mu^a \lambda_a 
- \sum_{a=0}^{8} \gamma^\mu \gamma^5 a_\mu^a \lambda_a \right) \Psi.
\label{lag_qm}
\end{align}
Here, $g_s$ and $g_v$ correspond to the scalar and vector coupling constants, $\Psi = (u,d,s)^T$ denotes the quark field, and $\bar{\Psi} = \Psi^\dagger \gamma^0$ is its Dirac adjoint, and $\Psi_{L,R}$ are the left- and right-handed components of the quark field. The quantities $\sigma_a$ and $\pi_a$ correspond to the scalar and pseudoscalar meson fields, and $v_\mu^a$ and $a_\mu^a$ denote the vector and axial-vector meson fields, respectively, and the Gell-Mann matrices, $\lambda_a$, are the generators of SU(3) flavor symmetry \cite{Kumari:2020mci}.
The scalar meson self-interaction Lagrangian responsible for spontaneous chiral symmetry breaking is expressed as \cite{Wang:2002pza} \begin{align}
\mathcal{L}_{\Sigma\Sigma} =\;
& -\frac{1}{2}k_{0}\chi^{2}\left(\sigma^{2} + \zeta^{2} + \delta^{2}\right)
+ k_{1}\left(\sigma^{2} + \zeta^{2} + \delta^{2}\right)^{2}
\nonumber \\
& + k_{2}\left(
\frac{\sigma^{4}}{2} + \frac{\delta^{4}}{2} + \zeta^{4}
+ 3\sigma^{2}\delta^{2}
\right)
+ k_{3}\chi\left(\sigma^{2} - \delta^{2}\right)\zeta
- k_{4}\chi^{4}
\nonumber \\
& - \frac{1}{4} \chi^4 \ln \frac{\chi^4}{\chi_0^4} + \frac{d}{3} \chi^4 \ln \left(\left(\frac{(\sigma^2-\delta^2)\zeta}{\sigma_0^2 \delta_0}\right)\left(\frac{\chi^3}{\chi_0^3}\right)\right).
\label{eq:L0}
\end{align}
In Table~\ref{tab:parameters}, the model parameters $k_{0}, k_{2}$ and $k_{4}$ are fitted to regenerate the vacuum conditions for the scalar fields ($\sigma,\zeta$) and dilaton field ($\chi$) and $k_{4}$ is fitted to reproduce the mass of $\sigma$ meson while the masses of $\eta$ and $\eta{'}$ fits $k_{3}$ parameter.
% fixed by reproducing vacuum hadron properties.
The scalar fields $\sigma$, $\zeta$, and $\delta$  couple to the light ($u, d$) and strange ($s$) quarks, encoding the effects of chiral symmetry breaking and the dilaton field $\chi$ is associated with the breaking of scale symmetry. The vector meson interaction Lagrangian is given by \cite{Wang:2002pza}
\begin{equation}
\mathcal{L}_{\mathrm{VV}} =
\frac{1}{2}\left(\frac{\chi}{\chi_{0}}\right)^2
\left( m_{\omega}^{2}\omega^{2} + m_{\rho}^{2}\rho^{2} + m_{\phi}^{2}\phi^{2} \right)
+ g_{4}
\left( \omega^{4} + 2\phi^{4} + 6\omega^{2}\rho^{2} + \rho^{4} \right).
\label{eq:Lvec}
\end{equation}
 In the above equation, $m_{\omega}$, $m_{\rho}$, and $m_{\phi}$ are the vacuum masses of the vector mesons where $\omega$, $\rho$  and $\phi$ are the vector fields that provide repulsive interactions. 
% The explicit breaking of scale invariance is incorporated through the dilaton potential,
% given by
% \begin{equation}
% \mathcal{L}_{\mathrm{scalebreak}} =
% -\frac{1}{4}\chi^{4}\ln\left(\frac{\chi^{4}}{\chi_{0}^{4}}\right)
% + \frac{d}{3}\chi^{4}
% \ln\left[
% \left(\frac{\sigma^{2}-\delta^{2}}{\sigma_{0}^{2}}\right)
% \left(\frac{\zeta}{\zeta_{0}}\right)
% \left(\frac{\chi}{\chi_{0}}\right)^{3}
% \right],
% \label{eq:Lscalebreak}
% \end{equation}
% where $\chi_{0}$, $\sigma_{0}$, and $\zeta_{0}$ denote the vacuum expectation values of
% the corresponding fields.
The explicit breaking of chiral symmetry due to non-zero current quark masses
is described by
\begin{equation}
\mathcal{L}_{\mathrm{\chi SB}} = -
\left(\frac{\chi}{\chi_{0}}\right)^{2}
\left[
m_{\pi}^{2}f_{\pi}\sigma
+ \left(\sqrt{2}m_{K}^{2}f_{K}
- \frac{1}{\sqrt{2}}m_{\pi}^{2}f_{\pi}\right)\zeta
\right].
\label{eq:LSB}
\end{equation}
The pion and kaon masses are denoted by $m_{\pi}$ and $m_{K}$, respectively, while $f_{\pi}$ and $f_{K}$
represent their corresponding leptonic decay constants \cite{Wang:2002pza}. The interaction of baryons to an external magnetic field is incorporated through the
electromagnetic interaction Lagrangian \cite{Mishra:2023uhx}
\begin{equation}
\mathcal{L}_{\mathrm{mag}} =
- \bar{\psi}_{i}\, q_{i}\gamma^{\mu}A_{\mu}\psi_{i}
- \frac{1}{2}\kappa_{i}\bar{\psi}_{i}\sigma^{\mu\nu}F_{\mu\nu}\psi_{i}
- \frac{1}{4}F^{\mu\nu}F_{\mu\nu}.
\label{l_mag}
\end{equation}
Here, $\bar\psi$ and $\psi$ represent wave functions of the $i^{th}$ baryon, $q_i$ and $\kappa_i$ denote the electric charge and the anomalous magnetic moment
of the $i^{th}$ baryon, respectively. The anomalous magnetic moment is defined as $ \kappa_i = \frac{\mu_i\, e_i}{2 M_p}$, where $\mu_i$ represents the intrinsic magnetic moment, expressed in units of nuclear magneton $\mu_N$, $e_i$ is the electric charge of the baryon and $M_p$ is the proton mass. Furthermore, $\gamma^\mu$ are the Dirac gamma matrices, the four-potential $A_\mu$ corresponds to the electromagnetic field, the associated field strength tensor is defined as $F_{\mu\nu}=\partial_{\mu}A_{\nu}-\partial_{\nu}A_{\mu}$, and
$\sigma^{\mu\nu}=\tfrac{i}{2}[\gamma^{\mu},\gamma^{\nu}]$ is the antisymmetric spin operator that governs the coupling between the baryon spin and the electromagnetic field \cite{Chakrabarty1997,Mukherjee:2018ebw}.
% The model parameters $k_{0}$, $k_{2}$, and $k_{4}$ are fixed by imposing vacuum extrema conditions on the scalar and dilaton fields, whereas $k_{1}$ is adjusted to reproduce the $\sigma$-meson mass and $k_{3}$ is constrained by the empirical $\eta$ and
% $\eta'$ meson masses. The vacuum value of the dilaton field $\chi$ is chosen such that the pressure vanishes at nuclear matter saturation density, ensuring thermodynamic
% consistency.
The total thermodynamic potential  is expressed as
\begin{equation}
\Omega = \Omega_{\mathrm{DS}} + \Omega_{\mathrm{med}}
- \mathcal{L}_M.
\label{eq:Omega_total}
\end{equation}
In above, contributions from
the Dirac sea (DS) and the thermal part (Fermi sea) of the baryons to the thermodynamic potential are denoted by $\Omega_{\mathrm{DS}}$ and $\Omega_{\mathrm{med}}$, respectively and $\mathcal{L}_M = \mathcal{L}_{\mathrm{\Sigma\Sigma}} + \mathcal{L}_{\mathrm{VV}} + \mathcal{L}_{\mathrm{\chi SB}}$ is the interaction between the mesons. These
contributions include both charged and neutral baryons, whose effective masses and
chemical potentials are modified through their interactions with the scalar and vector
mean fields. For spin-$\frac{1}{2}$ charged baryons ($p,\Sigma^{\pm},\Xi^{-}$), the DS
contribution to the thermodynamic potential is given by
\cite{Aguirre:2016vqa,Aguirre:2019ivr,Haber:2014ula}
\begin{equation}
\Omega^{\mathrm{charged}}_{\mathrm{DS}}
= \sum_{i}\Omega^{i}_{\mathrm{DS}}
= - \sum_{i}\frac{|q_i|B}{2\pi}
\left[
\sum_{\nu=0}^{\nu_{\max}}
\int_{0}^{\infty}\frac{dk_z}{2\pi}
\, \epsilon^{i}_{k,\nu,s=+1}
+
\sum_{\nu=1}^{\nu_{\max}}
\int_{0}^{\infty}\frac{dk_z}{2\pi}
\, \epsilon^{i}_{k,\nu,s=-1}
\right],
\end{equation} where $B$ denotes the
external magnetic field, $s$ labels the spin projection, and $\nu$ is the Landau level
index and $\nu_{max}$ represents the maximum value
of the Landau level. The thermal contribution of charged baryons to the thermodynamic potential is written as
 \cite{Strickland2012}
\begin{align}
\Omega^{\mathrm{charged}}_{\mathrm{med}}
&= \sum_{i}\Omega^{i}_{\mathrm{med}}
=-T \sum_{i}\frac{|q_i|B}{2\pi}
\Bigg[
\sum_{\nu=0}^{\nu_{\max}}
\int_{0}^{\infty}\frac{dk_z}{2\pi}
\Big\{
\ln\!\left(1+e^{-\beta(\epsilon^{i}_{k,\nu,s=+1}-\mu_i^{*})}\right)
\nonumber\\
&+ \ln\!\left(1+e^{-\beta(\epsilon^{i}_{k,\nu,s=+1}+\mu_i^{*})}\right)
\Big\}
+
\sum_{\nu=1}^{\nu_{\max}}
\int_{0}^{\infty}\frac{dk_z}{2\pi}
\Big\{
\ln\!\left(1+e^{-\beta(\epsilon^{i}_{k,\nu,s=-1}-\mu_i^{*})}\right)
\nonumber\\
&+ \ln\!\left(1+e^{-\beta(\epsilon^{i}_{k,\nu,s=-1}+\mu_i^{*})}\right)
\Big\}
\Bigg].
\end{align} Here, $\beta = 1/T$ is the inverse temperature and $\mu_i^{*}$ denotes the effective
chemical potential of the $i^{\mathrm{th}}$ charged baryon. The single-particle energy of a charged baryon in the presence of magnetic field is
given by
\begin{equation}
\epsilon^{i}_{k,\nu,s}
=
\sqrt{
k_z^{2}
+
\left(
\sqrt{2\nu|q_i|B + M_i^{*2}}
- s\,\kappa_i B
\right)^{2}
}.
\label{eq:energy_charged}
\end{equation}
The effective baryon mass is given by
\begin{equation}
M_i^{*} = \sqrt{E_i^{*2} - \langle p_{{i_{cm}}}^{2} \rangle},
\label{masses}
\end{equation}
and the term  $E_i^{*} = \sum_{q} n_{q_i}\, e_{q_i}^{*} + E_i^{\text{spin}}$ represents the  baryon energy, where $n_{q_i}$ is the number of quarks in the $i^{th}$ baryon with quark flavor q, and $e_{q_i}^{*}$ is the effective quark energy \cite{Shen:1999um}. The quantity $\langle p_{\mathrm{i_{cm}}}^{2} \rangle$ accounts for the correction due to the spurious center-of-mass motion \cite{Wang:2001jw,Barik1985}. 
% The self-consistent determination of the effective baryon mass $M_i^*$ establishes a connection between the underlying quark dynamics and the bulk properties of the system, such as pressure, energy density and binding energy.
% The effective mass of a quark of flavor $q$ ($u,d,s$) in the medium arises from its interaction
% with the scalar meson fields and is generated dynamically through spontaneous chiral
% symmetry breaking.
In the CQMF model framework, the in-medium
effective quark mass is obtained by the relation $m_q^{*} = - g_{\sigma}^q\,\sigma - g_{\zeta}^q\,\zeta - g_{\delta}^q\,\delta
\label{eq:quark_mass}$. Here, $g_{\sigma}^q$, $g_{\zeta}^q$ and $g_{\delta}^q$ denote the coupling constants
of the quark flavor, with scalar meson fields. The interaction of quarks with the vector meson fields leads to a shift in the chemical
potential. Accordingly, the effective chemical potential, accounting for its coupling to the vector mesons $\omega$, $\rho$ and
$\phi$, is expressed as $
\mu_q^{*} = \mu_q - g_{\omega}^q\,\omega - g_{\rho}^q\,\rho - g_{\phi}^q\,\phi ,
\label{eq:chemical_potential}
$ where $\mu_q^{*}$ is the effective quark chemical potential, $\mu_q$ is the quark chemical potential in the absence of vector interactions, $g_{\omega}^q$, $g_{\rho}^q$, and $g_{\phi}^q$ represent the corresponding vector coupling strengths.
% Following the formalism described above, we determine the effective masses of quarks and octet baryons in the presence of an external magnetic field. These masses serve as input for the calculation of magnetic moments, mentioned in the following sections. 
For neutral baryons ($n,\Lambda,\Sigma^{0},\Xi^{0}$), the DS contribution and thermal contribution to
the thermodynamic potential are given by \cite{Aguirre:2016vqa,Aguirre:2019ivr,Haber:2014ula}
\begin{equation}
\Omega^{\mathrm{neutral}}_{\mathrm{DS}}
= \sum_{i}\Omega^{i}_{\mathrm{DS}}
= - \sum_{i}\sum_{s=\pm 1}
\int \frac{d^{3}k}{(2\pi)^{3}}\, \epsilon^{i}_{k,s},
\label{eq:Omega_DS_neutral}
\end{equation}
and
\begin{align}
\Omega^{\mathrm{neutral}}_{\mathrm{med}}
&= \sum_{i}\Omega^{i}_{\mathrm{med}}
= -T \sum_{i}\sum_{s=\pm 1}
\int \frac{d^{3}k}{(2\pi)^{3}}
\left[
\ln\!\left(1+e^{-\beta(\epsilon^{i}_{k,s}-\mu_i^{*})}\right)
+ \ln\!\left(1+e^{-\beta(\epsilon^{i}_{k,s}+\mu_i^{*})}\right)
\right],
\label{eq:Omega_med_neutral}
\end{align} 
respectively. The single-particle energy of the $i^{\mathrm{th}}$ neutral baryon is given by
\begin{equation}
\epsilon^{i}_{k,s}
=
\sqrt{
k_z^{2}
+
\left(
\sqrt{k_x^{2}+k_y^{2}+M_i^{*2}}
- s\,\kappa_i B
\right)^{2}
}.
\label{eq:energy_neutral}
\end{equation}
% \begin{equation}
% M_B^* = \sum_{i=u,d,s} n_{iB} \, e_i^* - E_B^{(0)} + \Delta E_B,
% \end{equation}
% where $n_{iB}$ is the number of quarks of flavor $i$ in baryon $B$, $e_i^*$ is the effective quark energy given in Ref \cite{Singh:2016hiw}, and $\Delta E_B$ represents a confinement correction. The self-consistent generation of $M_B^*$ links microscopic quark interactions to macroscopic observables such as pressure, energy density, and binding energy.
% For a given baryon number density
% \begin{equation}
% \rho_B = \sum_{i}\rho_i,
% \end{equation}
% where $\rho_i$ is the number density of the $i^{\mathrm{th}}$ baryon, the isospin
% asymmetry parameter is defined as
% \begin{equation}
% \eta = -\frac{\sum_i I_{3i}\rho_i}{\rho_B},
% \end{equation}
% with $I_{3i}$ being the third component of isospin. The strangeness fraction is given by
% \begin{equation}
% f_s = \frac{\sum_i |s_i|\rho_i}{\rho_B},
% \end{equation}
% where $s_i$ denotes the number of strange quarks in the $i^{\mathrm{th}}$ baryon.
For fixed values of baryon density $\rho_B$, isospin asymmetry $I_a  = -\frac{\sum_{j} I^3_{j} \rho_{j}^{v}}{\rho_B}$, strangeness fraction $f_s = \frac{\sum_{j} |S_j| \rho_{j}^{v}}{\rho_B}$,  temperature $T$, and magnetic field $B$,
the equilibrium values of the scalar  and  vector fields are obtained by minimizing the
thermodynamic potential with respect to the corresponding fields \cite{Kumari:2020mci},
\begin{align}
\frac{\partial \Omega}{\partial \sigma} =
\frac{\partial \Omega}{\partial \zeta}  =
\frac{\partial \Omega}{\partial \delta} = 
\frac{\partial \Omega}{\partial \chi} = 
\frac{\partial \Omega}{\partial \omega} = 
\frac{\partial \Omega}{\partial \rho} = 
\frac{\partial \Omega}{\partial \phi} = 0.
\end{align}
This results in the following system of coupled equations:
\begin{align}
% &\frac{\partial \Omega}{\partial \sigma} =
&k_{0}\chi^{2}\sigma
- 4k_{1}(\sigma^{2}+\zeta^{2}+\delta^{2})\sigma
- 2k_{2}(\sigma^{3}+3\sigma\delta^{2})
- 2k_{3}\chi\sigma\zeta 
- \frac{d}{3}\chi^{4}\left(\frac{2\sigma}{\sigma^{2}-\delta^{2}}\right) && \nonumber \\
&
+ \left(\frac{\chi}{\chi_{0}}\right)^{2} m_{\pi}^{2}f_{\pi}
- \left(\frac{\chi}{\chi_{0}}\right)^{2} m_{\omega}\omega^{2}
\frac{\partial m_{\omega}}{\partial \sigma}
- \left(\frac{\chi}{\chi_{0}}\right)^{2} m_{\rho}\rho^{2}
\frac{\partial m_{\rho}}{\partial \sigma}
- \sum_{i} g_{\sigma_i}\rho_{i}^{s}
= 0, &&
\label{eq:sigma_field}
\end{align}

\begin{align}
% &\frac{\partial \Omega}{\partial \zeta} =
&k_{0}\chi^{2}\zeta
- 4k_{1}(\sigma^{2}+\zeta^{2}+\delta^{2})\zeta
- 4k_{2}\zeta^{3}
- k_{3}\chi(\sigma^{2}-\delta^{2})
- \frac{d}{3}\frac{\chi^{4}}{\zeta} + \left(\frac{\chi}{\chi_{0}}\right)^{2}
 && \nonumber \\
&
\left(\sqrt{2}m_{K}^{2}f_{K} - \frac{1}{\sqrt{2}}m_{\pi}^{2}f_{\pi}
\right) - \left(\frac{\chi}{\chi_{0}}\right)^{2} m_{\phi}\phi^{2}
\frac{\partial m_{\phi}}{\partial \zeta}
- \sum_{i} g_{\zeta_i}\rho_{i}^{s}
= 0, &&
\label{eq:zeta_field}
\end{align}

\begin{align}
&
k_{0}\chi^{2}\delta
- 4k_{1}(\sigma^{2}+\zeta^{2}+\delta^{2})\delta
- 2k_{2}(\delta^{3}+3\sigma^{2}\delta) + 2k_{3}\chi\delta\zeta
+ \frac{2}{3}d\chi^{4}
\left(\frac{\delta}{\sigma^{2}-\delta^{2}}\right) - \sum_{i} g_{\delta_i}\rho_{i}^{s} = 0, &&
\label{eq:delta_field}
\end{align}

\begin{align}
&
k_{0}\chi(\sigma^{2}+\zeta^{2}+\delta^{2})
- k_{3}(\sigma^{2}-\delta^{2})\zeta
+ \chi^{3}\left[1+\ln\left(\frac{\chi^{4}}{\chi_{0}^{4}}\right)\right]
+ (4k_{4}-d)\chi^{3} - \frac{4}{3}d\chi^{3} && \nonumber \\
&
\ln\!\left[
\left(\frac{\sigma^{2}-\delta^{2}}{\sigma_{0}^{2}}\right)
\left(\frac{\zeta}{\zeta_{0}}\right)
\left(\frac{\chi}{\chi_{0}}\right)^{3}
\right] + \frac{2\chi}{\chi_{0}^{2}}
\left[
m_{\pi}^{2}f_{\pi}\sigma
+ \left(\sqrt{2}m_{K}^{2}f_{K}
- \frac{1}{\sqrt{2}}m_{\pi}^{2}f_{\pi}\right)\zeta
\right] && \nonumber \\
&
% && \nonumber \\
% &\quad
- \frac{\chi}{\chi_{0}^{2}}
\left(
m_{\omega}^{2}\omega^{2}
+ m_{\rho}^{2}\rho^{2}
+ m_{\phi}^{2}\phi^{2}
\right)
= 0, &&
\label{eq:chi_field}
\end{align}

\begin{align}
&
\frac{\chi^{2}}{\chi_{0}^{2}} m_{\omega}^{2}\omega
+ 4g_{4}\omega^{3}
+ 12g_{4}\omega\rho^{2}
- \sum_{i} g_{\omega_i}\rho_{i}^{v}
= 0, &&
\label{eq:omega_field}
\end{align}

\begin{align}
&
\frac{\chi^{2}}{\chi_{0}^{2}} m_{\rho}^{2}\rho
+ 4g_{4}\rho^{3}
+ 12g_{4}\omega^{2}\rho
- \sum_{i} g_{\rho_i}\rho_{i}^{v}
= 0, &&
\label{eq:rho_field}
\end{align}

\begin{align}
&
\frac{\chi^{2}}{\chi_{0}^{2}} m_{\phi}^{2}\phi
+ 8g_{4}\phi^{3}
- \sum_{i} g_{\phi_i}\rho_{i}^{v}
= 0. &&
\label{eq:phi_field}
\end{align}

\renewcommand{\thetable}{\arabic{table}}
%table
\begin{table}[htbp]
\centering
\renewcommand{\arraystretch}{1.25}
\setlength{\tabcolsep}{6pt}
\resizebox{\textwidth}{!}
{
\begin{tabular}{V c|c|c|c|c|c|c|c|c V}
% \begin{tabular}{!{\vrule width 1.5pt}c|c|c|c|c|c|c|c|c!{\vrule width 1.5pt}}

\noalign{\hrule height 1.5pt}

$k_{0}$ & $k_{1}$ & $k_{2}$ & $k_{3}$ & $k_{4}$ & $g_{s}$ & $g_{v}$ & $g_{4}$ & $\rho_{0}$ (fm$^{-3}$)\\
\hline
4.94 & 2.12 & $-10.16$ & $-5.38$ & $-0.06$ & 3.85 & 9.14 & 37.5 & 0.16 \\
\hline
$\sigma_{0}$ (MeV) & $\zeta_{0}$ (MeV) & $\chi_{0}$ (MeV)  & $f_{\pi}$ (MeV) & $m_\pi$ (MeV) & $m_{K}$ (MeV) & $f_{K}$ (MeV) & $m_{\omega}$ (MeV) & $m_{\phi}$ (MeV) \\
\hline
$-93$ & $-96.87$ & 254.38 & 93 & 139 & 496 & 115 & 783 & 1020 \\
\hline
$g_{\sigma}^{u}$ & $g_{\sigma}^{d}$ & $g_{\sigma}^{s}$ & $g_{\zeta}^{u}$ & $g_{\zeta}^{d}$ & $g_{\zeta}^{s}$ & $g_{\delta}^{u}$ & $g_{\delta}^{d}$ & $g_{\delta}^{s}$ \\
\hline
2.72 & 2.72 & 0 & 0 & 0 & 3.85 & 2.72 & 2.72 & 0 \\
\hline
$g_{\omega}^{u}$ & $g_{\omega}^{d}$ & $g_{\omega}^{s}$ & $g_{\phi}^{u}$ & $g_{\phi}^{d}$ & $g_{\phi}^{s}$ & $g_{\rho}^{u}$ & $g_{\rho}^{d}$ & $g_{\rho}^{s}$ \\
\hline

% \hline
% $g_{\omega}^{u}$ & $g_{\omega}^{d}$ & $g_{\omega}^{s}$ & $g_{\phi}^{u}$ & $g_{\phi}^{d}$ & $g_{\phi}^{s}$ & $g_{\rho}^{u}$ & $g_{\rho}^{d}$ & $g_{\rho}^{s}$ \\
% \hline

3.23 & 3.23 & 0 & 0 & 0 & 4.57 & 3.23 & 3.23 & 0 \\

\hline
$g_{\sigma_p}$ & $g_{\zeta _p}$ & $g_{\delta _p}$ & $g_{\sigma_n}$ & $g_{\zeta _n}$ & $g_{\delta _n}$ & $g_{\sigma _\Lambda}$ & $g_{\zeta_\Lambda}$ & $g_{\delta_\Lambda}$ \\
\hline
6.64 & 0 & 2.72 & 6.64 & 0 & 2.72 & 5.44 & 3.85 & 0 \\
\hline

$g_{\sigma_\Sigma}$ & $g_{\zeta_\Sigma}$ & $g_{\delta_\Sigma}$ & $g_{\sigma_\Xi}$ & $g_{\zeta_\Xi}$ & $g_{\delta_\Xi}$ & $g_{\omega_p}$ & $g_{\rho_p}$ & $g_{\phi_p}$ \\
\hline
5.44 & 3.85 & 2.72 & 2.72 & 7.69 & 2.72 & 9.69 & 8.89 & 0 \\
\hline

$g_{\omega_n}$ & $g_{\rho_n}$ & $g_{\phi_n}$ & $g_{\omega_\Lambda}$ & $g_{\rho_\Lambda}$ & $g_{\phi_\Lambda}$ & $g_{\omega_\Sigma}$ & $g_{\rho_\Sigma}$ & $g_{\phi_\Sigma}$ \\
\hline
9.69 & 8.89 & 0 & 8.85 & 0 & $-4.57$ & 12.28 & 12.28 & $-4.57$ \\
\hline

$g_{\omega_\Xi}$ & $g_{\rho_\Xi}$ & $g_{\phi_\Xi}$ & -- & -- & -- & -- & -- & -- \\
\hline
7.27 & 3.23 & $-9.14$ & -- & -- & -- & -- & -- & -- \\

\noalign{\hrule height 1.5pt}

\end{tabular}
}
\caption{List of values of parameters used in the present work \cite{Wang:2002pza}.}
\label{tab:parameters}
\end{table}
The derivative terms involving meson masses with respect to scalar fields, such as $\frac{\partial m_{\omega}}{\partial \sigma}$, $\frac{\partial m_{\rho}}{\partial \sigma}$, and $\frac{\partial m_{\phi}}{\partial \zeta}$ appearing in the Eqs.~\eqref{eq:sigma_field}, \eqref{eq:zeta_field} arises due to the implicit dependence of meson masses on the scalar fields within the chiral SU(3) framework. The parameters employed in solving the coupled field equations are listed in Table~\ref{tab:parameters} \cite{Wang:2002pza,Kumari:2020mci}.
The number density of baryons ($i =p,n,\Lambda,\Sigma^{0},\Sigma^{\pm},\Xi^{0},\Xi^{-}$) is obtained from the
thermal contributions of the thermodynamic potential. In the presence of external magnetic field, the expression for scalar and vector densities differ for charged and neutral baryons due to Landau quantization effects
\cite{Chakrabarty1997,Kumar:2019axp,Chahal:2024oib,Broderick:2000pe,Broderick:2000pe}. For the $i^{th}$ charged baryons, the number density is given by
\begin{align}
\rho_{i}^{v}
&=
\frac{|q_i|B}{2\pi^{2}}
\Bigg[
\sum_{\nu=0}^{\nu_{\max}}
\int_{0}^{\infty} dk_z
\left(
\frac{1}{1+e^{\beta(\epsilon^{i}_{k,\nu,s=+1}-\mu_i^{*})}}
-
\frac{1}{1+e^{\beta(\epsilon^{i}_{k,\nu,s=+1}+\mu_i^{*})}}
\right)
\nonumber \\
&\quad
+
\sum_{\nu=1}^{\nu_{\max}}
\int_{0}^{\infty} dk_z
\left(
\frac{1}{1+e^{\beta(\epsilon^{i}_{k,\nu,s=-1}-\mu_i^{*})}}
-
\frac{1}{1+e^{\beta(\epsilon^{i}_{k,\nu,s=-1}+\mu_i^{*})}}
\right)
\Bigg].
\label{eq:number_density_charged_fixed}
\end{align}  
For the $i^{th}$ neutral baryons, which do not undergo Landau quantization, the number density
reduces to
\begin{equation}
\rho_{i}^{v} =
\sum_{s=\pm1}
\int \frac{d^{3}k}{(2\pi)^{3}}
\left(
\frac{1}{1+e^{\beta(\epsilon^{i}_{k,s}-\mu_i^{*})}}
-
\frac{1}{1+e^{\beta(\epsilon^{i}_{k,s}+\mu_i^{*})}}
\right).
\label{eq:number_density_neutral}
\end{equation}
The scalar density of the $i^{\mathrm{th}}$ baryon is defined as
$\rho_i^{s}=\langle \bar{\psi}_i \psi_i \rangle
= \partial\Omega/\partial M_i^{*}$.
Since the scalar fields $(\sigma,\zeta,\delta)$ explicitly depend on the scalar
densities through the effective baryon masses, the field equations are solved
self-consistently \cite{Broderick:2001qw,Broderick:2000pe}. In the presence of magnetic field, the scalar density
for charged baryons is given by
% \begin{align}
% \rho_{i}^{s}
% &=
% \frac{|q_i|B}{2\pi^{2}} M_i^*
% \Bigg[
% \sum_{\nu=0}^{\nu_{\max}}
% \int_{0}^{\infty} d k_z
% \frac{\sqrt{{M_i^*}^2+2\nu |q_i| B}-s\kappa_i B}{\epsilon^{i}_{k,\nu,s=+1}
% \sqrt{{M_i^*}^2+2\nu |q_i| B}}
% \left(
% \frac{1}{1+e^{\beta(\epsilon^{i}_{k,\nu,s=+1}-\mu_i^{*})}}
% -
% \frac{1}{1+e^{\beta(\epsilon^{i}_{k,\nu,s=+1}+\mu_i^{*})}}
% \right)
% \nonumber \\
% \nonumber \\
% &\quad
% +
% \sum_{\nu=1}^{\nu_{\max}}
% \int_{0}^{\infty} d k_z
% \frac{\sqrt{{M_i^*}^2+2\nu |q_i| B} - s\kappa_i B}{\epsilon^{i}_{k,\nu,s=-1}
% \sqrt{{M_i^*}^2+2\nu |q_i| B}}
% \left(
% \frac{1}{1+e^{\beta(\epsilon^{i}_{k,\nu,s=-1}-\mu_i^{*})}}
% -
% \frac{1}{1+e^{\beta(\epsilon^{i}_{k,\nu,s=-1}+\mu_i^{*})}}
% \right)
% \Bigg].
% \label{eq:scalar_density_charged_fixed}
% \end{align}

\begin{align}
\rho_{i}^{s}
&=
\frac{|q_i|B}{2\pi^{2}} M_i^*
\Bigg[
\sum_{\nu=0}^{\nu_{\max}}
\int_{0}^{\infty} d k_z\,
\frac{\sqrt{{M_i^*}^2+2\nu |q_i| B}-s\kappa_i B}
{\epsilon^{i}_{k,\nu,s=+1}
\sqrt{{M_i^*}^2+2\nu |q_i| B}}
\nonumber\\[2mm]
&\qquad\times
\left(
\frac{1}{1+e^{\beta(\epsilon^{i}_{k,\nu,s=+1}-\mu_i^{*})}}
-
\frac{1}{1+e^{\beta(\epsilon^{i}_{k,\nu,s=+1}+\mu_i^{*})}}
\right)
\nonumber\\[4mm]
&\qquad+
\sum_{\nu=1}^{\nu_{\max}}
\int_{0}^{\infty} d k_z\,
\frac{\sqrt{{M_i^*}^2+2\nu |q_i| B}-s\kappa_i B}
{\epsilon^{i}_{k,\nu,s=-1}
\sqrt{{M_i^*}^2+2\nu |q_i| B}}
\nonumber\\[2mm]
&\qquad\times
\left(
\frac{1}{1+e^{\beta(\epsilon^{i}_{k,\nu,s=-1}-\mu_i^{*})}}
-
\frac{1}{1+e^{\beta(\epsilon^{i}_{k,\nu,s=-1}+\mu_i^{*})}}
\right)
\Bigg].
\label{eq:scalar_density_charged_fixed}
\end{align}
% \begin{align}
% \rho_s^i  \equiv  {\rho_s^i}^{med} &=\frac{|q_i| B}{2\pi^2} 
% \sum_{s = \pm 1}\sum_{\nu=0}^{\infty} (2-\delta_{\nu 0}) 
% m_i^* \int_{0}^\infty d k_z
% \frac{\sqrt{{m_i^*}^2+2\nu |q_i| B}-s\kappa_i B}{\epsilon^{i}_{k,\nu,s}
% \sqrt{{m_i^*}^2+2\nu |q_i| B}}
% \nonumber \\ &
% \times \Bigg [
% \frac {1}{1+e^{\beta (\epsilon^{i}_{k,\nu, s}-\mu^{*}_i)}}
% +\frac {1}{1+e^{\beta (\epsilon^{i}_{k,\nu, s}+\mu^{*}_i)}}
% \Bigg],
% \label{rho_s_i_ch_med}
% \end{align}
For neutral baryons, the scalar density is expressed as
\begin{equation}
\rho_i^{s}
=
M_i^{*}
\sum_{s=\pm1}
\int \frac{d^{3}k}{(2\pi)^{3}}
\frac{\sqrt{k_x^{2}+k_y^{2}+M_i^{*2}}-s\kappa_i B}
{\epsilon^{i}_{k,s}\sqrt{k_x^{2}+k_y^{2}+M_i^{*2}}}
\left(
\frac{1}{1+e^{\beta(\epsilon^{i}_{k,s}-\mu_i^{*})}}
+
\frac{1}{1+e^{\beta(\epsilon^{i}_{k,s}+\mu_i^{*})}}
\right).
\label{eq:scalar_density_neutral}
\end{equation}
The contribution of the magnetized DS to the scalar density of the
$i^{\mathrm{th}}$ baryon is given by
\cite{Haber:2014ula,Mukherjee:2018ebw}
\begin{equation}
\rho^{\mathrm{DS},i}_{s}
=
-\frac{1}{4\pi^{2}}
\left[
\frac{(q_i B)^{2}}{3M_i^{*}}
+
\left(
(\kappa_i B)^{2} M_i^{*}
+
(|q_i|B)(\,\kappa_i B)
\left(
\frac{1}{2}
+
2\ln\!\left(\frac{M_i^{*}}{M_i}\right)
\right) \right)
\right].
\label{eq:scalar_density_DS}
\end{equation}
%The DS density is derived from the energy equation for DS \cite{Mishra:2023uhx,Haber:2014ula,Mukherjee:2018ebw}.
 The effects of the DS are incorporated by adding the contribution $\rho^{\mathrm{DS},i}_{s}$ to the total scalar density of each baryon species.
The expectation values of the fields are obtained by solving the corresponding coupled equations of motion independently. These equations are solved in the presence of an external magnetic field for a fixed temperature ($T$ = 100 MeV), which lies in the hot hadronic regime, where thermal effects and partial restoration of chiral symmetry become significant while the system remains predominantly in the confined hadronic phase. The calculations are performed at different baryon densities ($\rho_B = 0, \rho_0, \rho_0/2$) and in asymmetric strange matter.
% For finite baryon density, the system is characterized by the total baryon
% density $\rho_{B}$, the isospin asymmetry parameter $\eta$, and the strangeness
% fraction $f_{s}$. The self-consistent solutions of the mean-field equations
% determine the in-medium masses of hadrons embedded in the magnetized medium.
% The scalar and vector densities entering these equations are given by
% \begin{align}
% \rho_{s,B} &= \frac{\gamma_B}{(2\pi)^3} \int \frac{M_B^*}{\sqrt{k^2 + M_B^{*2}}} \, n_B(k) \, d^3k, \\
% \rho_B &= \frac{\gamma_B}{(2\pi)^3} \int n_B(k) \, d^3k,
% \end{align}
% where $n_B(k)$ is the Fermi–Dirac distribution. The inclusion of hyperons such as $\Lambda$, $\Sigma$, and $\Xi$ ensures that both non-strange and strange degrees of freedom are consistently treated.
% \subsubsection*{4. Scale Breaking and the Dilaton Field}
% The scale invariance of QCD is broken through the introduction of a scalar dilaton field $\chi$. It reproduces the trace anomaly and ensures proper scale breaking consistent with QCD phenomenology. []
% , whose potential is expressed as
% \begin{equation}
% V_\chi = \frac{1}{4} B \left( \frac{\chi}{\chi_0} \right)^4 
% \left( \ln\frac{\chi^4}{\chi_0^4} - 1 \right),
% \end{equation}
% where $B$ is the vacuum energy density and $\chi_0$ is the vacuum expectation value. This term reproduces the trace anomaly and ensures proper scale breaking consistent with QCD phenomenology.
The CQMF model effectively captures the evolution of quark and baryon masses in dense, strange, and asymmetric matter. It provides the framework to analyze how scalar and vector meson condensates influence hadronic properties, strangeness content, and the overall equation of state. When magnetic fields are included, the model accounts for Landau quantization and charge dependent magnetic effects. This makes it suitable for studying magnetized strange matter as well as application to compact stellar systems.

\subsection*{B. The Chiral Constituent Quark Model}

The $\chi$CQM is employed alongside the CQMF model to describe the internal quark structure 
of baryons, including contributions from valence quarks, polarized quark sea, and orbital angular momentum. In this framework, the simple constituent quark picture is extended through chiral fluctuations involving the emission of Goldstone bosons associated with 
spontaneous chiral symmetry breaking ~\cite{Dahiya:2009ix}. The model has been 
successful in describing the magnetic moments of octet baryons. Within the $\chi$CQM framework, the total magnetic moment of a baryon is written 
as the sum of valence, sea, and orbital contributions \cite{Manohar:1983md,Cheng:1997tt,Gupta2003}, i.e.,
\begin{equation}
\mu_{total}^* = \mu_{\text{val}}^* + \mu_{\text{sea}}^* + \mu_{\text{orb}}^*.
\end{equation}
Configuration mixing affects the spin structure of the baryon by modifying the 
probabilities of different quark spin alignments, leading to improved agreement 
with experimental observations. In addition, the quark-Goldstone boson fluctuation 
process,
$
q^\uparrow \rightarrow GB + q'^\downarrow  ,
$
introduces polarized sea quark that contribute to the total magnetic moment \cite{Manohar:1983md}.

% \subsubsection*{2. Sea Quark Polarization and Goldstone Boson Emission}

% The emission probabilities of different Goldstone bosons ($\pi$, $K$, $\eta$, $\eta'$) are determined by SU(3) breaking parameters $\alpha$, $\beta$, and $\zeta$, which control the relative strength of transitions between flavors. The resulting spin polarizations of the sea quarks are obtained by considering the coupling of quarks to the pseudoscalar meson cloud, producing both spin-flip and orbital contributions.

% \subsubsection*{3. Integration with the CQMF Model}

When combined with the CQMF model, the $\chi$CQM provides a consistent way to calculate the magnetic moments of baryons in the medium. The effective quark and baryon masses obtained from the CQMF equations are used as input to compute the in-medium magnetic moments. Thus, the hybrid CQMF–$\chi$CQM framework links the dynamical mass generation of quarks to their spin and orbital contributions within baryons and orbital angular momentum of sea quarks, respectively \cite{Cheng:1997tt,Cheng:1997kr}. The effective magnetic moment of valence and sea quarks are given by \cite{Dahiya:2023izc}
\begin{equation}
\mu_{\mathrm{val}}^{*} = \sum_{q=u,d,s,c} \Delta q_{\mathrm{val}}\, \mu_q^{*}; \quad\quad
\mu_{\mathrm{sea}}^{*} = \sum_{q=u,d,s,c} \Delta q_{\mathrm{sea}}\, \mu_q^{*},
\label{eq:mu_val_sea}
\end{equation}
where \(\Delta q_{\mathrm{val}}\) and 
\(\Delta q_{\mathrm{sea}}\) account for spin polarization, incorporating configuration mixing. To evaluate the medium-modified magnetic moments of quarks, we 
employ the following relations \cite{Dahiya:2023izc}:
\begin{equation}
\mu_d^{*} = - \frac{1}{2}\mu_u^{*} = 
-\left( 1 - \frac{\Delta M}{M_i^{*}} \right), \quad
\mu_s^{*} = - \frac{m_u^{*}}{m_s^{*}}
\left( 1 - \frac{\Delta M}{M_i^{*}} \right), \quad
\mu_c^{*} = \frac{2 m_u^{*}}{m_c}
\left( 1 - \frac{\Delta M}{M_i^{*}} \right).
\label{eq:quark_magnetic}
\end{equation}
The quark magnetic moment expressions employed here are constructed so as to 
account for both confinement effects and relativistic corrections in a consistent 
manner while evaluating the magnetic moments of baryons. In this framework, these 
quantities are commonly interpreted as the mass-adjusted magnetic moments of the
constituent quarks \cite{Girdhar:2015gsa,Dutt:2024lui}. The in-medium modification 
of the baryon mass is characterized by the mass shift 
$\Delta M = M_{\mathrm{vac}} - M_i^{*}$. Here, $M_{\mathrm{vac}}$ denotes the 
vacuum baryon mass \cite{Singh:2016hiw}. The contribution to the magnetic moment arising from the orbital motion of sea 
quarks is given by
\begin{equation}
\mu_{\mathrm{orb}}^{*} =
\sum_{q=u,d,s,c} \Delta q_{\mathrm{val}}\,
\mu^{*}(q_{+} \rightarrow q'_{-}).
\label{eq:mu_orbit}
\end{equation}
In the above relation, $\mu^{*}(q_{+} \rightarrow q'_{-})$ represents the orbital magnetic moment 
corresponding to an individual chiral fluctuation \cite{Gupta2003}. These terms are 
weighted according to the probabilities associated with the respective chiral 
transitions \cite{Sharma2010}.

\begin{equation}
\begin{aligned}
\mu^*(u_{+} \rightarrow u_-) =\; a \Bigg[
& \frac{3{m_u^{*}}^{2}}{2m_{\pi}({m_u^{*}} + m_{\pi})}
- \frac{\alpha^{2}(m_{K}^{2} - 3{m_u^{*}}^{2})}{2m_{K}({m_u^{*}} + m_{K})}
+ \frac{\gamma^{2} M_D}{({m_u^{*}} + M_D)} \\
& + \frac{\beta^{2} M_{\eta}}{6({m_u^{*}} + M_{\eta})}
+ \frac{\zeta^{2} M_{\eta'}}{48({m_u^{*}} + M_{\eta'})}
+ \frac{\gamma^{2} M_{\eta_c}}{16({m_u^{*}} + M_{\eta_c})}
\Bigg] \mu_N ,
\end{aligned}
\label{eq:u+}
\end{equation}

\begin{equation}
\begin{aligned}
\mu^*(d_{+} \rightarrow d_-) =\; a \frac{m_u^{*}}{m_d^{*}} \Bigg[
& \frac{3(m_{\pi}^{2} - 2{m_d^{*}}^{2})}{4m_{\pi}({m_d^{*2}} + m_{\pi})}
- \frac{\alpha^{2} m_{K}}{2({m_d^{*}} + m_{K})}
+ \frac{\gamma^{2}(2M_{D}^{2} - 3{m_d^{*}}^{2})}{2M_{D}({m_d^{*}} + M_D)} \\
& - \frac{\beta^{2} M_{\eta}}{12({m_d^{*}} + M_{\eta})}
- \frac{\zeta^{2} M_{\eta'}}{96({m_d^{*}} + M_{\eta'})}
+ \frac{\gamma^{2} M_{\eta_c}}{32({m_d^{*}} + M_{D})}
\Bigg] \mu_N ,
\end{aligned}
\label{eq:d+}
\end{equation}

\begin{equation}
\begin{aligned}
\mu^*(s_{+} \rightarrow s_-) =\; a \frac{m_u^{*}}{m_s^{*}} \Bigg[
& \frac{\alpha^{2}(m_{K}^{2} - 3{m_s^{*}}^{2})}{2m_{K}({m_s^{*}} + m_{K})}
+ \frac{\gamma^{2}(2M_{D_s}^{2} - 3{m_s^{*}}^{2})}{2M_{D_s}({m_s^{*}} + M_{D_s})} \\
& - \frac{\beta^{2} M_{\eta}}{3({m_s^{*}} + M_{\eta})}
- \frac{\zeta^{2} M_{\eta'}}{96({m_s^{*}} + M_{\eta'})}
- \frac{\gamma^{2} M_{\eta_c}}{32({m_s^{*}} + M_{\eta_c})}
\Bigg] \mu_N ,
\end{aligned}
\label{eq:s+}
\end{equation}

% \vspace{-1cm}
\begin{equation}
\begin{aligned}
\mu^*(c_{+} \rightarrow c_-) =\; a \frac{m_u^{*}}{m_c} \Bigg[
& \frac{\gamma^{2}(M_{D}^{2} + 3m_c^{2})}{2M_{D}(m_c + M_{D}^2)}
- \frac{\gamma^{2}(M_{D_s}^{2} + 3m_c^{2})}{2M_{D_s}(m_c + M_{D_s}^2)} + \frac{3\zeta^{2} M_{\eta'}}{16(m_c + M_{\eta'})} \\
& 
+ \frac{9\gamma^{2} M_{\eta_c}}{16(m_c + M_{\eta_c})}
\Bigg] \mu_N.
\end{aligned}
\label{eq:c+}
\end{equation}
The description of all the terms is given in Ref. \cite{Dutt:2024lui}. The values of meson masses, $M_{\eta}$, $M_{\eta'}$, $M_{\eta_c}$, $M_D$ and $M_{D_s}$ mentioned in Table~\ref{tab:my_constants}   \cite{ParticleDataGroup:2024cfk}.
\begin{table}[htbp]
\centering
\renewcommand{\arraystretch}{1.25}
\setlength{\tabcolsep}{6pt}
\resizebox{\textwidth}{!}{%
\begin{tabular}{V c|c|c|c|c|c V}

\noalign{\hrule height 1.5pt}

$a$ & $a{\alpha}^2$ & $a{\beta}^2$ & $a{\zeta}^2$ & $a{\gamma}^2$  & $m_c$ (MeV) \\
\hline
0.12 & 0.0243 & 0.0243 & 0.0053 & 0.0014 & 1270 \\
\hline
$M_{\eta}$ (MeV) & $M_{\eta'}$ (MeV) & $M_{\eta_c}$ (MeV) & $M_D$ (MeV) & $M_{D_s}$ (MeV)  & - \\
\hline
547.862 & 957.78 & 2983.5 & 1869 & 1968 & - \\
\noalign{\hrule height 1.2pt}
\end{tabular}%
}
\caption{Constants used in the present work to calculate the magnetic moments containing the meson masses and the fitting parameters \cite{ParticleDataGroup:2024cfk}.}
\label{tab:my_constants}
\end{table}
% \subsubsection*{4. Physical Significance}
The $\chi$CQM describes the baryon magnetic moments which come from the contribution of the valence, sea, and orbital parts, while the CQMF model determines how these quantities evolve under external conditions. The combined framework allows the study of magnetic moment modifications due to density, strangeness, and magnetic field effects, offering insights into the spin structure and magnetization of strange hadronic matter. Together, they provide a unified description connecting chiral dynamics, quark mass evolution, and baryonic magnetic properties in extreme environments.

% \clearpage
% \FloatBarrier
\section{\label{Results}Results and Discussions}

In this section, we analyze the results obtained within the CQMF-$\chi$CQM framework. The discussion is divided into two parts. In subsection \ref{Eq_subsection_scalar_1}, we discuss the behavior of scalar meson fields and effective masses of octet baryons in the presence of external magnetic field. 
The results on the magnetic moments of octet baryons in the magnetized strange matter are given in subsection \ref{Sec_magentic_moment1}.

\subsection{Scalar Meson Fields and Effective Masses of Octet Baryons}
\label{Eq_subsection_scalar_1}
In this section, we examine the behavior of the scalar fields and effective masses of octet baryons as functions of the external magnetic field. Previous studies have shown that, in the absence of DS contributions, the fields and effective baryon masses exhibit only weak dependence on the external magnetic field at zero density \cite{Mishra:2023uhx}. This
behavior originates from the mechanism of mass generation in the chiral mean field
framework, where baryon masses are primarily governed by the scalar fields $\sigma$ and $\zeta$, which remain close to their vacuum values at zero density. In the absence of density driven feedback and vacuum polarization effects, the external magnetic fields do not induce significant modifications of the values scalar condensates. Even though charged baryons experience Landau quantization, their contribution to bulk properties remains weak when DS effects are not included.
% the lack of a Dirac sea at $T=100$ MeV prevents Landau-level effects from contributing appreciably to the bulk properties.
Consequently, a weak variation of the effective masses is observed in charged and neutral baryons.
% \begin{figure}[t]
% \centering

% \begin{subfigure}{0.24\textwidth}
% \includegraphics[width=\linewidth]{sigma_f_vs_BB.png}
% \caption{$\sigma,\ \eta=0$}
% \end{subfigure}
% \hfill
% \begin{subfigure}{0.24\textwidth}
% \includegraphics[width=\linewidth]{sigma_f_vs_BB_eta_0p5.png}
% \caption{$\sigma,\ \eta=0.5$}
% \end{subfigure}
% \hfill
% \begin{subfigure}{0.24\textwidth}
% \includegraphics[width=\linewidth]{zeta_f_vs_BB.png}
% \caption{$\zeta,\ \eta=0$}
% \end{subfigure}
% \hfill
% \begin{subfigure}{0.24\textwidth}
% \includegraphics[width=\linewidth]{zeta_f_vs_BB_eta_0p5.png}
% \caption{$\zeta,\ \eta=0.5$}

% \end{subfigure}

% \caption{
% Magnetic field dependence of the scalar fields $\sigma$ and $\zeta$ 
% at $T=100$ MeV for isospin symmetric ($\eta=0$) 
% and asymmetric ($\eta=0.5$) matter.
% }
% \label{fig:scalar_fields}
% \end{figure}
\begin{figure}[htbp]
\centering
\vspace{-1.2cm}
% Row 1
\begin{subfigure}{0.42\textwidth}
\centering
% \vspace{-1cm}
\captionsetup{justification=centering, labelformat=empty}
\caption{\boldmath$I_a = 0$}
\includegraphics[width=\linewidth]{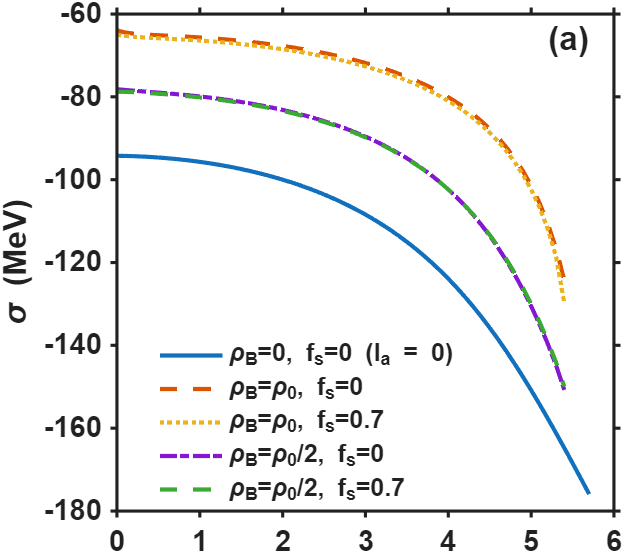}
\end{subfigure}
\hfill
\begin{subfigure}{0.42\textwidth}
\centering
% \vspace{-1cm}
\captionsetup{justification=centering, labelformat=empty}
\caption{\boldmath$I_a = 0.5$}
\includegraphics[width=\linewidth]{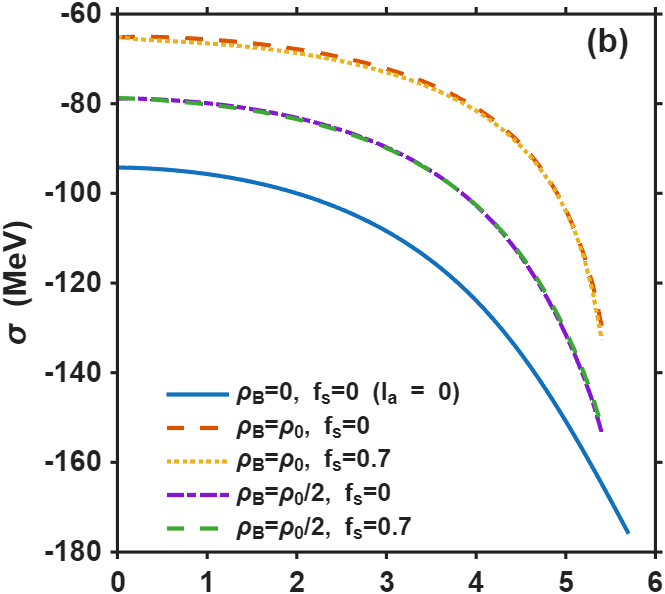}
\end{subfigure}

\vspace{0.5cm}

% Row 2
\begin{subfigure}{0.42\textwidth}
\centering

\includegraphics[width=\linewidth]{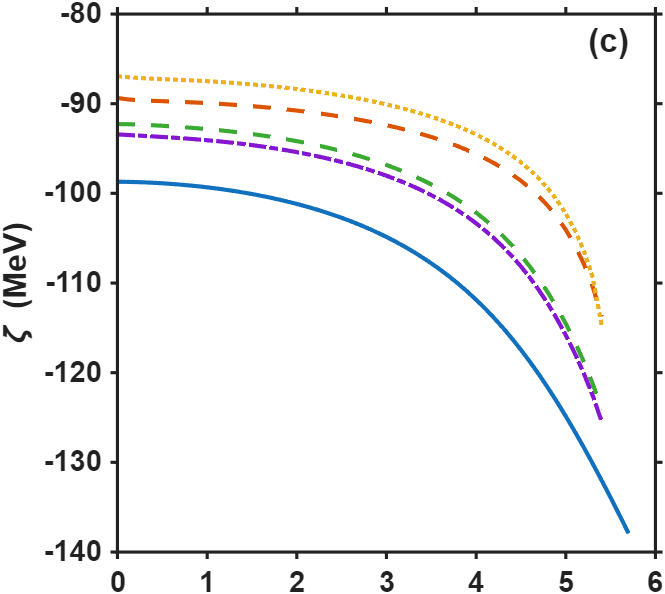}
% \caption{$\zeta,\ \eta=0$}
\end{subfigure}
\hfill
\begin{subfigure}{0.42\textwidth}
\centering
\includegraphics[width=\linewidth]{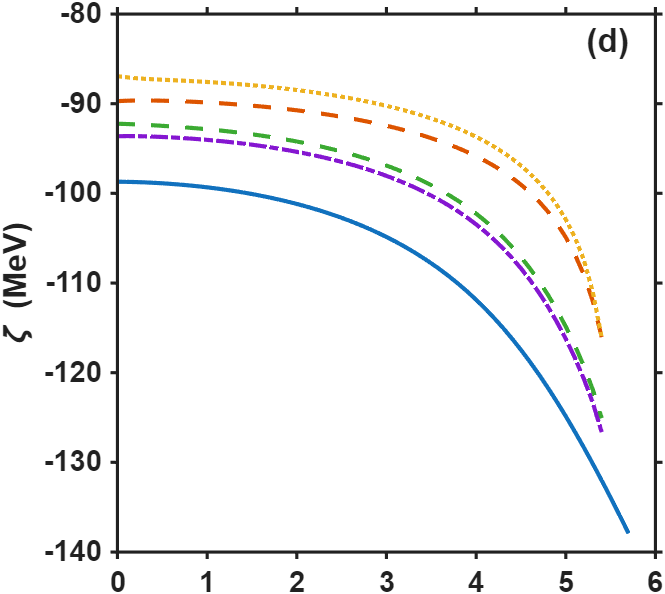}
% \caption{$\zeta,\ \eta=0.5$}
% \label{fig:zeta_fields}
\end{subfigure}

\vspace{0.5cm}

% Row 3
\begin{subfigure}{0.42\textwidth}
\centering
\includegraphics[width=\linewidth]{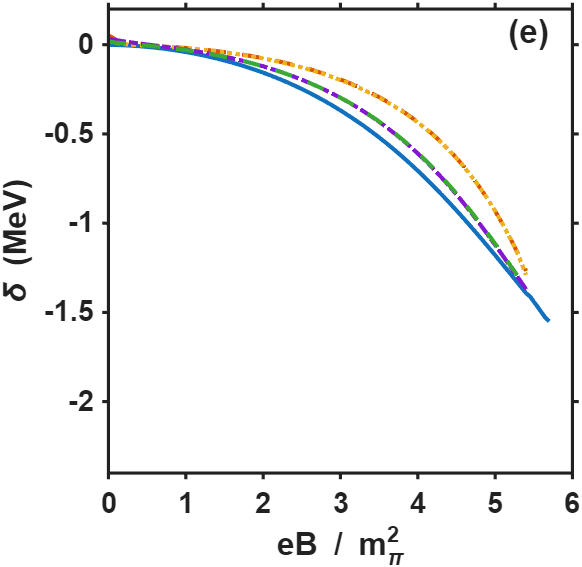}

\end{subfigure}
\hfill
\begin{subfigure}{0.42\textwidth}
\centering
\includegraphics[width=\linewidth]{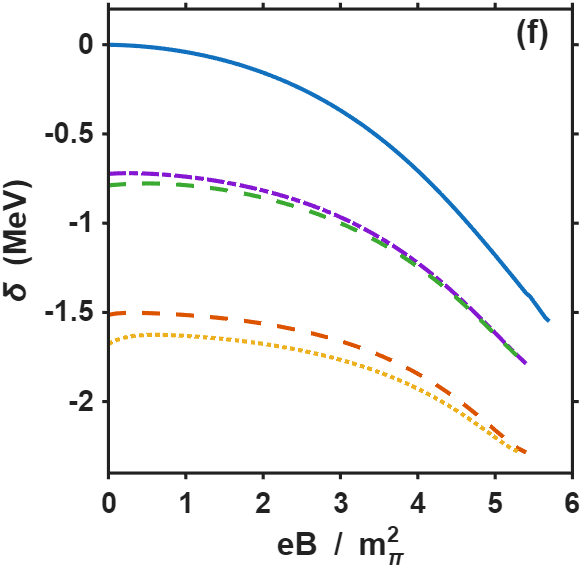}

% \label{fig:scalar_fields}
\end{subfigure}

\caption{ Dependence of the scalar fields $\sigma$, $\zeta$, and $\delta$ on external magnetic field  
at $T=100$ MeV. Subplots (a), (c), and (e) correspond to $I_a=0$, while subplots (b), (d), and (f) represent $I_a=0.5$. In each subplot, results are shown for baryon densities, $\rho_B=0$, $\rho_0/2$ and $\rho_0$, and strangeness fraction, $f_s=0$ and 0.7 and are further compared with the case, $\rho_B = 0$, $I_a=0$ and $f_s=0$.
}

\label{fig:scalar_fields}

\end{figure}

In Fig.~\ref{fig:scalar_fields}, we present the variation of the $\sigma$, $\zeta$ and $\delta$ scalar fields as functions of the external magnetic field $eB/m_\pi^2$ at temperature $T=100$ MeV within the CQMF model including Landau quantization and the magnetic field dependent DS contribution for vacuum ($\rho_B=0$) and finite baryon densities ($\rho_0/2, \rho_0$), at different $I_a$ and $f_s$ values.
The magnitude of the $\sigma$ field increases monotonically with 
increasing DS magnetic field.
% i.e., $\sigma$ becomes progressively 
% more negative.
This behavior reflects magnetic catalysis induced by 
the DS contribution, where Landau quantization of charged 
baryons enhances the scalar density and consequently strengthens 
the chiral condensates and effect becomes particularly pronounced 
for magnetic fields higher than $eB \gtrsim 3\,m_\pi^2$. Comparing Figs.~\ref{fig:scalar_fields} (a) and (b), the enhancement due to magnetic field is persistent across all densities, 
indicating that magnetic catalysis dominates over density effects 
at sufficiently large $eB$. The presence of $f_s$ has only a negligible effect on the fields. Although isospin asymmetry ($I_a = 0.5$) modifies the magnitude of the $\sigma$ field, the overall monotonic trend and rise with magnetic field remain largely unaffected since $\sigma$ is a non-strange field.

A qualitatively similar monotonic behavior is observed for $\zeta$ in Figs.~\ref{fig:scalar_fields} (c) and (d), where it also becomes more negative with increasing magnetic field, indicating strengthening of the strange quark condensate. However, the magnetic-field dependence of $\zeta$ is comparatively weaker than that of $\sigma$. This relatively weak sensitivity can originate from 
the larger strange quark mass and its weaker coupling to magnetic field effects. At finite density, the magnitude of $\zeta$ shifts toward smaller negative values compared to the vacuum, indicating partial modification to the strange condensate. However, with increasing magnetic field strength, the condensate exhibits an overall increment across all densities and matter compositions. However, with increasing magnetic field strength, the condensate shows an overall increase across all densities and compositions. The inclusion of strangeness fraction mainly shifts the magnitude of $\zeta$, without significantly affecting its monotonic behavior.

In contrast, the scalar-isovector field $\delta$, shown in Figs.~\ref{fig:scalar_fields} (e) and (f), exhibits a much stronger sensitivity to both isospin asymmetry and magnetic field 
effects. Even for isospin symmetric matter ($I_a=0$), the $\delta$ field does not remain exactly zero but develops a small negative value, which increases in magnitude 
with the magnetic field. This behavior arises due to Landau quantization \cite{Rabhi:2011ej}, which affects charged baryons differently depending on their electric charge, leading 
to an effective isospin imbalance even for $I_a = 0$. For asymmetric matter ($I_a = 0.5$), the $\delta$ field is already finite at zero magnetic field due to the finite isospin asymmetry of the system. As the magnetic field strength increases, the $\delta$ field becomes increasingly negative, indicating variation in isospin splitting . The variation is more clear compared to the symmetric case, reflecting the combined effect of initial asymmetry and magnetic field contributions. In both cases, the overall trend is governed by the interplay between density, isospin asymmetry, and Landau quantization, while the effect of strangeness fraction remains subdominant, leading only to moderate changes in the field magnitudes.

The present results are consistent with earlier studies of magnetized hadronic and quark matter. In relativistic mean-field approaches such as the Walecka model, it has been shown that the magnetic-field dependence of nucleon masses becomes significant only when DS (vacuum polarization) effects are included~\cite{Mukherjee:2018ebw}. In agreement with these findings, the scalar fields in the present work exhibit an increasing magnitude with magnetic field, reflecting the role of Landau quantization and its impact on the scalar condensates. A similar trend is observed in NJL model calculations, where the phenomenon of magnetic catalysis leads to rise of the quark condensate with increasing magnetic field ~\cite{Klevansky:1992qe,Miransky:2015ava}. The variation in the $\sigma$, $\zeta$ and $\delta$ fields obtained here follows the same underlying mechanism. However, unlike the NJL model, the CQMF framework treats baryons as bound states of quarks and incorporates confinement effects, making it more suitable for describing dense hadronic matter at finite density.

The weaker sensitivity of the $\zeta$ field to the magnetic field is also in line with previous chiral hadronic model where the strange sector shows a reduced response due to the larger strange quark mass and weaker coupling~\cite{Papazoglou1999}. Furthermore, the reduction of scalar fields at finite density, followed by their variation in the presence of a magnetic field, agrees with earlier results from chiral SU(3) and quark-meson coupling models~\cite{Saito:2005rv}. Since the present calculations are performed at $T = 100$ MeV, which lies below the chiral symmetry restoration temperature, the observed behavior remains consistent with magnetic catalysis reported in effective models and lattice QCD studies~\cite{Bali2012}. At higher temperatures, approaching the chiral restoration region, thermal effects are expected to weaken the scalar condensates, which may lead to a reduction in the increase of the magnitude of scalar fields caused due to increased magnetic field strength~\cite{Borsanyi:2020fev, Farias:2021fci}.

\begin{figure}[htbp]
\centering
% \renewcommand{\thefigure}{4}
% \vspace{0.5cm}
\begin{subfigure}{0.42\textwidth}
\centering

\captionsetup{justification=centering,labelformat=empty}
\caption{\boldmath$I_a=0$}
\includegraphics[width=\linewidth]{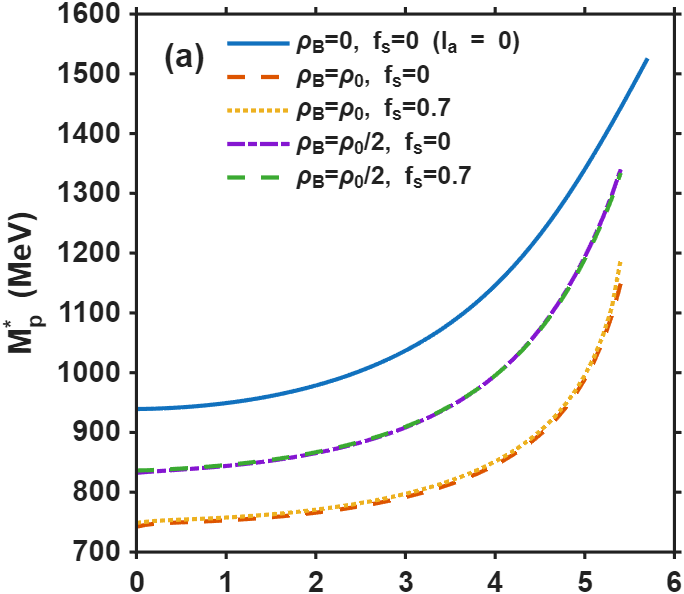}
% \caption{$p,\ \rho=\rho_0/2$}
\end{subfigure}
\hfill
\begin{subfigure}{0.42\textwidth}
\centering
\captionsetup{justification=centering,labelformat=empty}
\caption{\boldmath$I_a=0.5$}
\includegraphics[width=\linewidth]{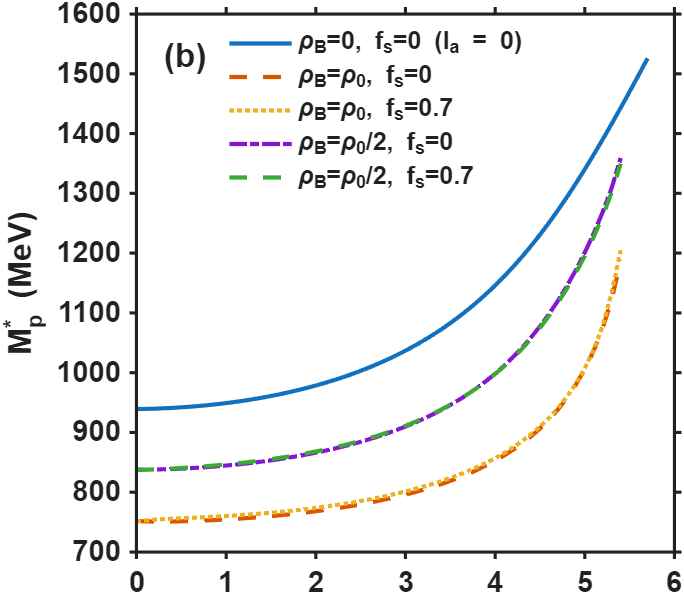}
% \caption{$p,\ \rho=\rho_0$}
\end{subfigure}

\vspace{0.5cm}

\begin{subfigure}{0.42\textwidth}
\centering
\includegraphics[width=\linewidth]{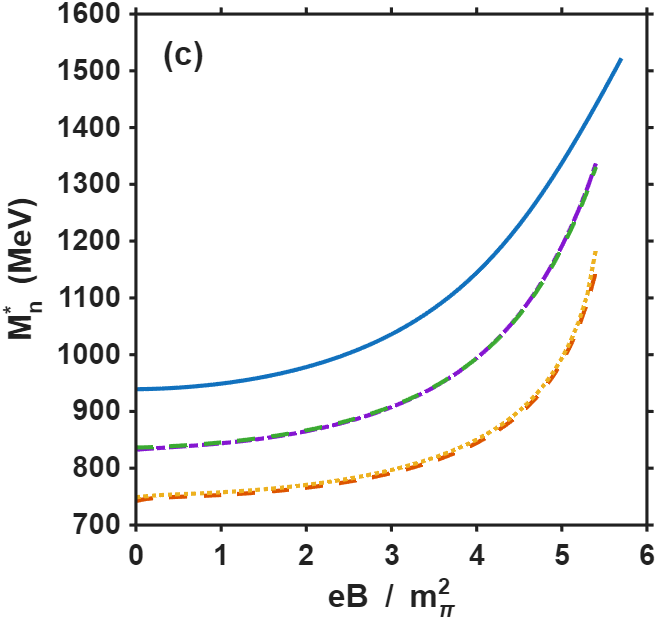}

\end{subfigure}
\hfill
\begin{subfigure}{0.42\textwidth}
\centering
\includegraphics[width=\linewidth]{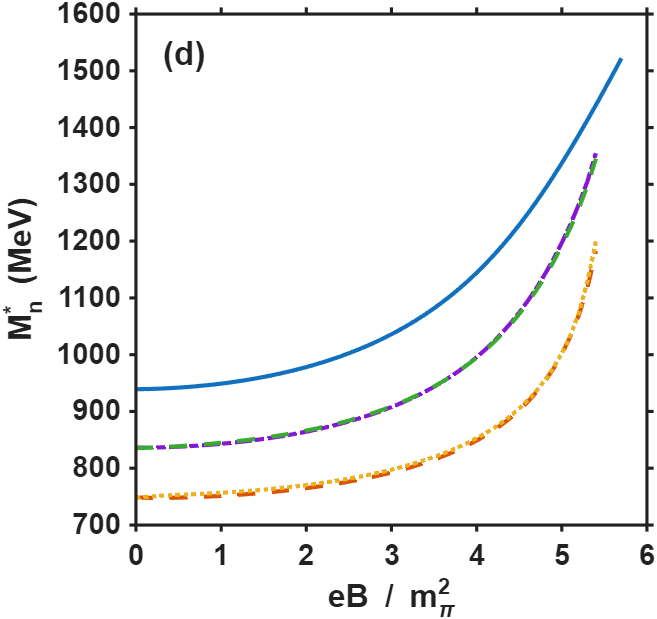}
% \label{fig:nucleon_effective_masses}
\end{subfigure}

\caption{Dependence of the effective masses of nucleons ($p$, $n$) on external magnetic field at $T=100$ MeV. Subplots (a) and (c) correspond to $I_a=0$, while subplots
(b) and (d) represent $I_a=0.5$. In each subplot, results are shown for baryon densities, 
$\rho_B=0$,$\rho_0/2$ and $\rho_0$, and strangeness fraction, $f_s=0$ and 0.7, and are further also compared with the case, $\rho_B = 0$, $I_a=0$ and $f_s=0$.}
\label{fig:nucleon_effective_masses}
\end{figure}
We now discuss the magnetic field dependence of the effective masses, given by Eq.~\ref{masses}, of nucleons shown in Figs.~\ref{fig:nucleon_effective_masses} (a), (b) for the proton and (c), (d) for the neutron corresponding to symmetric ($I_a = 0$) and asymmetric ($I_a = 0.5$) matter at baryon densities $\rho_B = 0$, $\rho_0/2$, and $\rho_0$, and strangeness fractions $f_s = 0$ and $0.7$. In all cases, the effective masses increase monotonically with magnetic field strength. The variation remains relatively weak in the low magnetic-field region ($eB \lesssim 2-3\,m_\pi^2$), while a much stronger nonlinear rise is observed beyond $eB \sim 3\,m_\pi^2$, as clearly reflected in both Fig.~\ref{fig:nucleon_effective_masses} and Table~\ref{tab:masses}. For the proton in vacuum ($\rho_B=0$, $f_s=0$), the effective mass increases from $939.16$ MeV at $eB=0$ to about $1029$ MeV at $eB=3\,m_\pi^2$.
% , corresponding to an enhancement of nearly $10\%$. 
With further increase in magnetic field, the mass rises sharply to $1316.59$ MeV at $eB=5\,m_\pi^2$,
% giving an overall increase of about $40\%$,
as listed in Table~\ref{tab:masses}. A similar trend is observed for the neutron, whose effective mass increases from $939.16$ MeV at $eB=0$ to $1313.85$ MeV at $eB=5\,m_\pi^2$. The modest variation at low magnetic fields followed by a rapid increment at higher $eB$ indicates that the scalar condensates become increasingly sensitive to magnetic field effects in the strongly magnetized regime. The proton exhibits a slightly stronger magnetic field response than the neutron due to its direct coupling to the external magnetic field through Landau quantization \cite{Broderick:2000pe}. 
%add the below in paper%
% At low magnetic fields, the contribution from the lowest Landau levels remains small, leading to comparatively mild changes in the effective masses. However, as the magnetic field strength increases, the Landau-level spacing becomes larger and significantly modifies the scalar densities, thereby producing a rapid enhancement in the proton mass. In contrast, the neutron does not undergo Landau quantization because of its neutral charge, and its magnetic-field dependence arises mainly through anomalous magnetic moment interactions and medium-induced scalar-field modifications \cite{Broderick:2000pe}. 
Consequently, at $eB \sim 5\,m_\pi^2$, the proton mass remains slightly larger than the neutron mass by a few MeV for all considered medium configurations.

\begin{figure}[htbp]
\centering
\vspace{-1cm}

% Sigma0
\begin{subfigure}{0.42\textwidth}
\centering
\captionsetup{justification=centering,labelformat=empty}
\caption{\boldmath$I_a=0$}
\includegraphics[width=\linewidth]{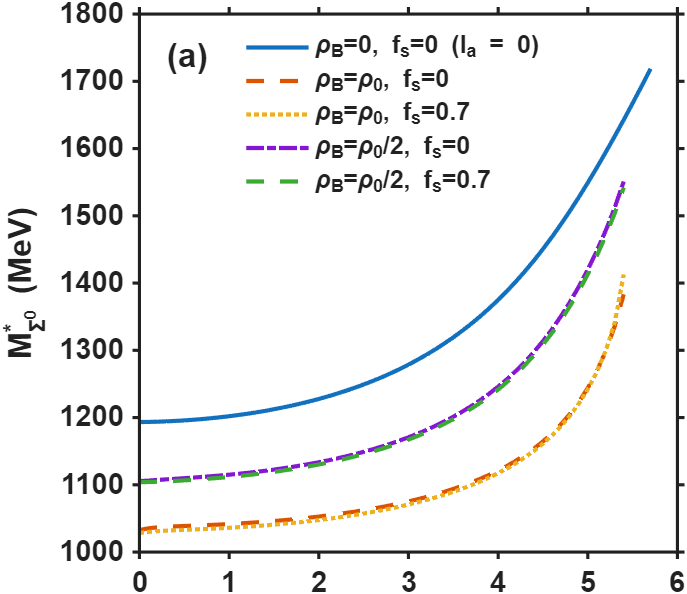}

\end{subfigure}
\hfill
\begin{subfigure}{0.42\textwidth}
\centering
\captionsetup{justification=centering,labelformat=empty}
\caption{\boldmath$I_a=0.5$}
\includegraphics[width=\linewidth]{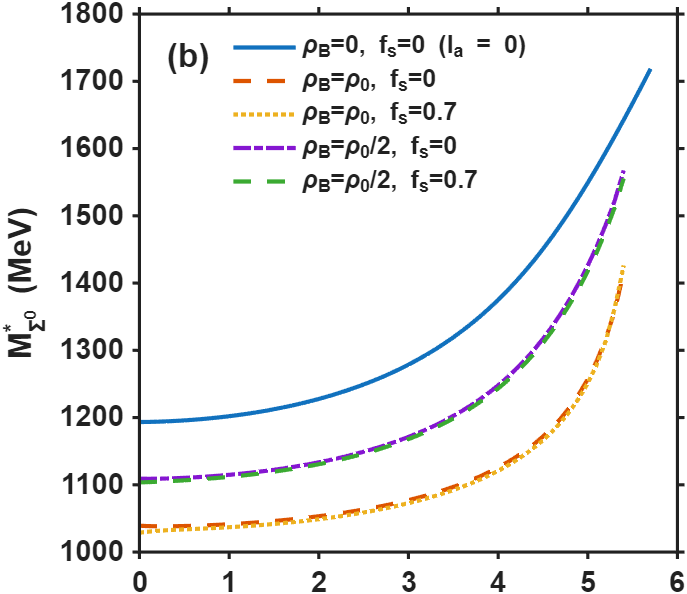}

\end{subfigure}
\vspace{0.5cm}

% Sigma+
\begin{subfigure}{0.42\textwidth}
\centering
\includegraphics[width=\linewidth]{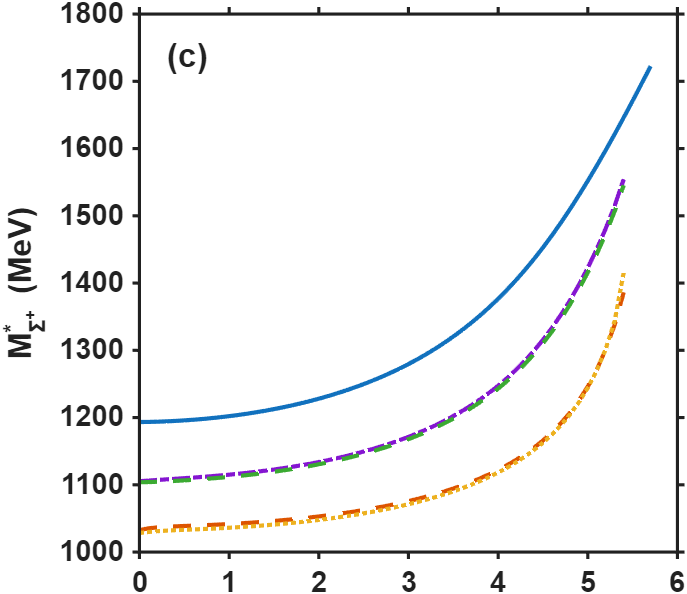}

\end{subfigure}
\hfill
\begin{subfigure}{0.42\textwidth}
\centering
\includegraphics[width=\linewidth]{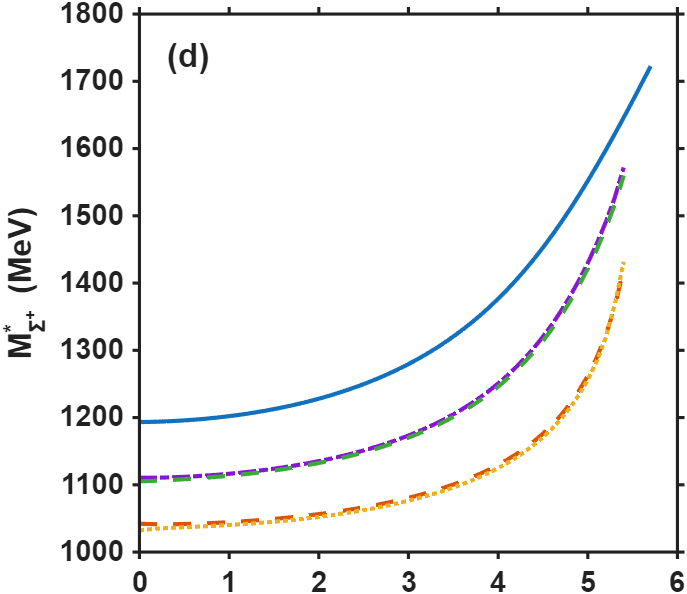}

\end{subfigure}

\vspace{0.5cm}

\begin{subfigure}{0.42\textwidth}
\centering
% \captionsetup{justification=centering,labelformat=empty}
% \caption{\boldmath$I_a=0$}
\includegraphics[width=\linewidth]{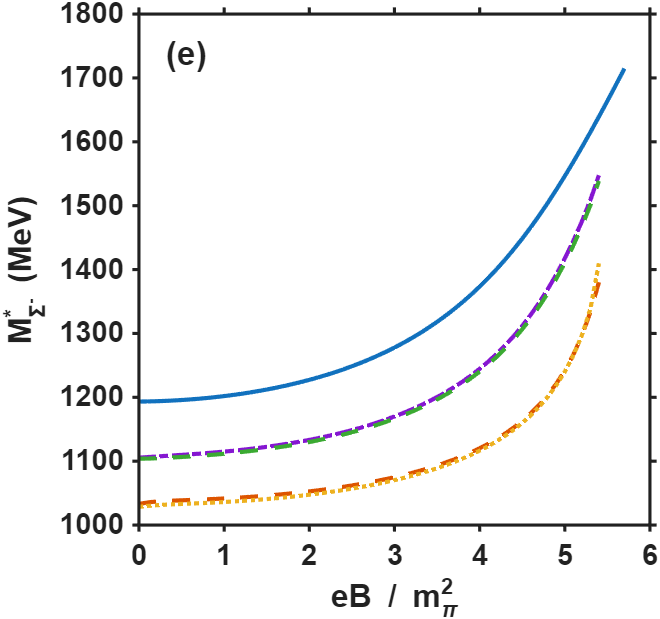}
\end{subfigure}
\hfill
\begin{subfigure}{0.42\textwidth}
\centering
% \captionsetup{justification=centering,labelformat=empty}
% \caption{\boldmath$I_a=0.5$}
\includegraphics[width=\linewidth]{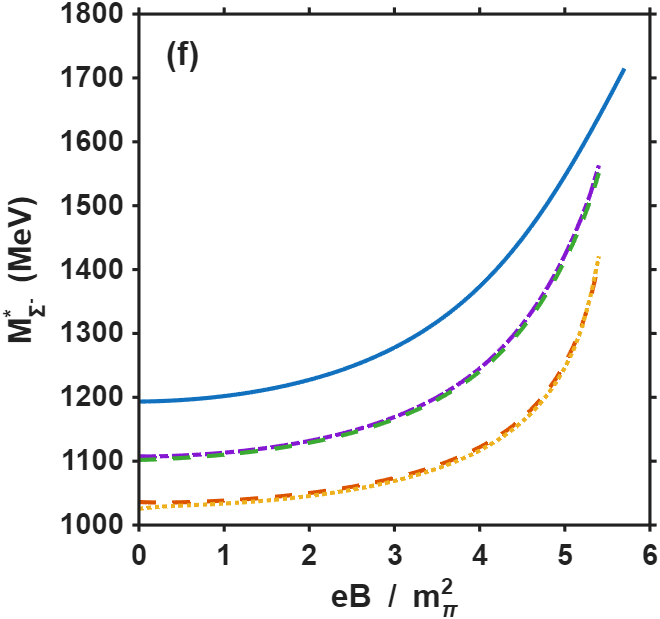}
\end{subfigure}

\vspace{0.2cm}

\caption{Dependence of the effective masses ($M_i^*$) of hyperons ($\Sigma^0$, $\Sigma^+$, $\Sigma^-$) on the external magnetic field at $T=100$ MeV. Subplots (a), (c), and (e) correspond to $I_a=0$, while subplots (b), (d), and (f) represent $I_a=0.5$. In each subplot, results are shown for baryon densities, $\rho_B=0$, $\rho_0/2$ and $\rho_0$, and strangeness fraction, $f_s=0$ and 0.7, and are also compared with the case, $\rho_B = 0$, $I_a=0$ and $f_s=0$.}
\label{fig:hyperon_1_effective_masses}
\end{figure}

\begin{figure}[htbp]
\centering
% \renewcommand{\thefigure}{3.2}
% \vspace{-2cm}
% Sigma-
\vspace{-1cm}
% Lambda
\begin{subfigure}{0.42\textwidth}
\centering
\captionsetup{justification=centering,labelformat=empty}
\caption{\boldmath$I_a=0$}

\includegraphics[width=\linewidth]{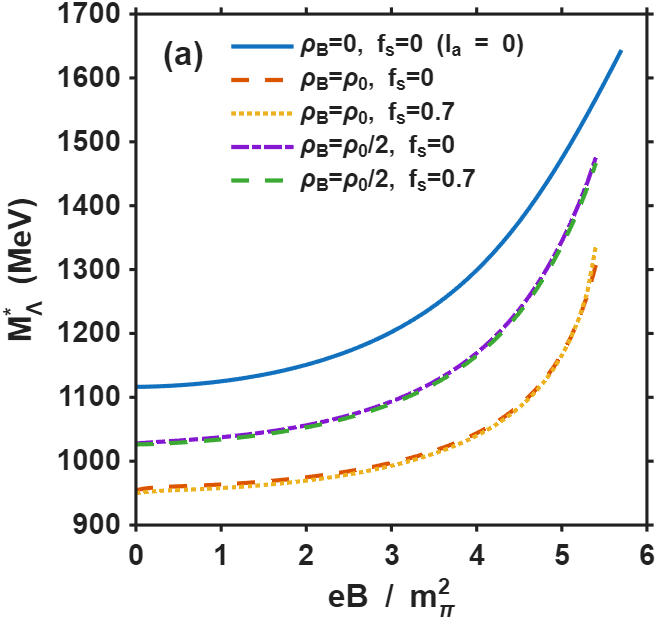}
\end{subfigure}
\hfill
\begin{subfigure}{0.42\textwidth}
\centering
\captionsetup{justification=centering,labelformat=empty}
\caption{\boldmath$I_a=0.5$}
\includegraphics[width=\linewidth]{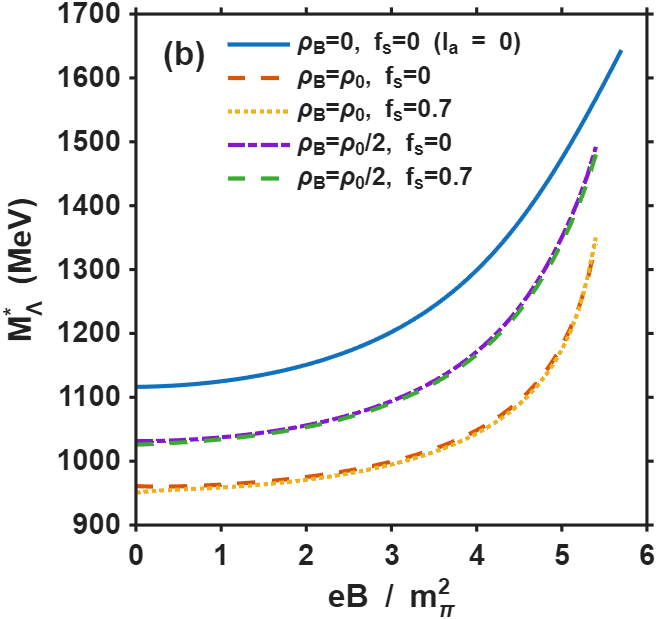}
\end{subfigure}

\caption{Magnetic field dependence of the effective masses ($M_i^*$) of hyperon ($\Lambda$) at $T=100$ MeV. Subplot (a) corresponds to $I_a=0$, while subplot (b) represents $I_a=0.5$. In each subplot, results are shown for baryon densities, $\rho_B=0$, $\rho_0/2$ and $\rho_0$, and strangeness fraction, $f_s=0$ and 0.7, and are further compared with the case, $\rho_B = 0$, $I_a=0$ and $f_s=0$.}
\label{Lambda_masses}
\end{figure}

\begin{figure}[htbp]
\centering
% \renewcommand{\thefigure}{3.2}
% \vspace{-2cm}
% Sigma-
% \vspace{-1.5cm}
% Lambda

% \vspace{0.5cm}

% Xi0
\begin{subfigure}{0.42\textwidth}
\centering
\captionsetup{justification=centering,labelformat=empty}
\caption{\boldmath$I_a=0$}
\includegraphics[width=\linewidth]{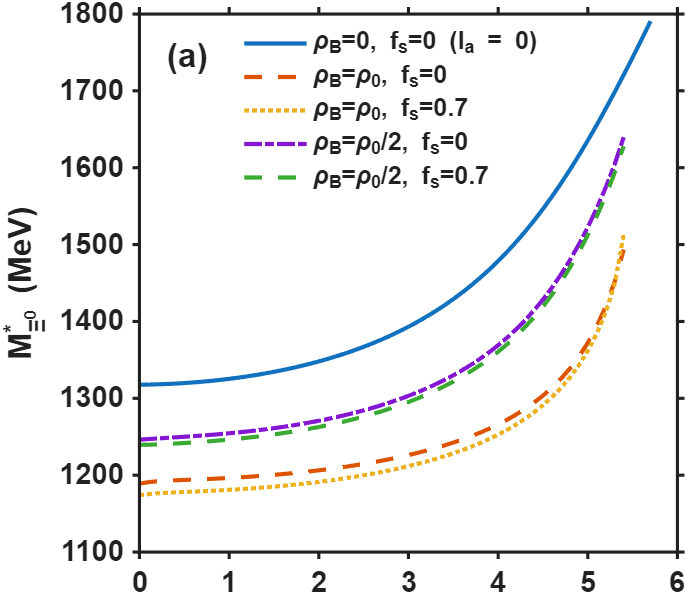}
\end{subfigure}
\hfill
\begin{subfigure}{0.42\textwidth}
\centering
\captionsetup{justification=centering,labelformat=empty}
\caption{\boldmath$I_a=0.5$}
\includegraphics[width=\linewidth]{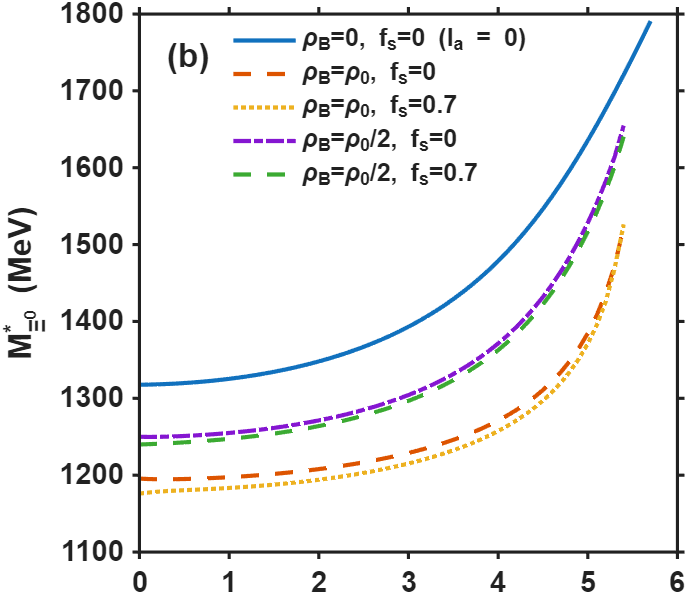}
\end{subfigure}

\vspace{0.5cm}

% Xi-
\begin{subfigure}{0.42\textwidth}
\centering
\includegraphics[width=\linewidth]{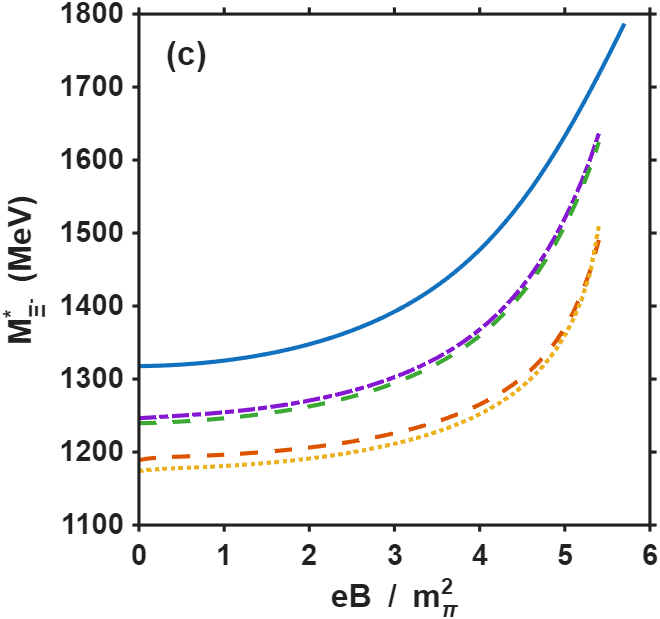}

\end{subfigure}
\hfill
\begin{subfigure}{0.42\textwidth}
\centering
\includegraphics[width=\linewidth]{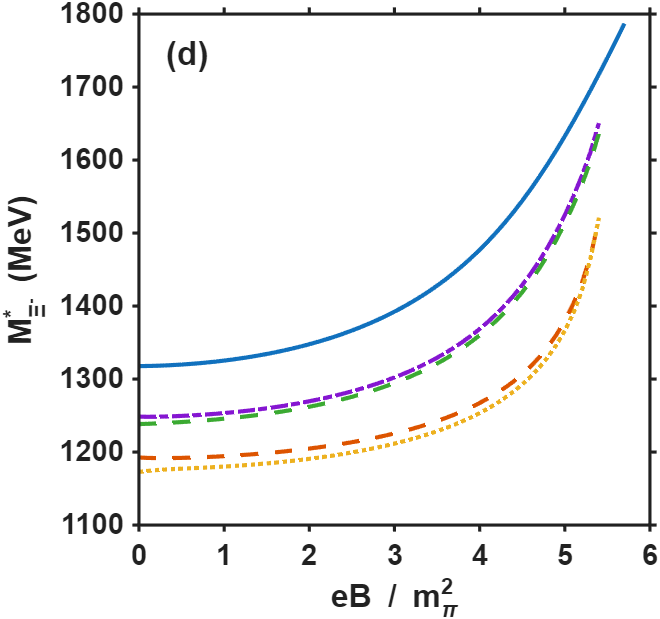}

\end{subfigure}

\caption{Magnetic field dependence of the effective masses ($M_i^*$) of hyperons ($\Xi^0$, $\Xi^-$) at $T=100$ MeV. Subplots (a) and (c) correspond to $I_a=0$, while subplots (b) and (d) represent $I_a=0.5$. In each subplot, results are shown for baryon densities, $\rho_B=0$, $\rho_0/2$ and $\rho_0$, and strangeness fraction, $f_s=0$ and 0.7, and are also compared with the case, $\rho_B = 0$, $I_a=0$ and $f_s=0$.}
\label{fig:hyperon_2_effective_masses}
\end{figure}

At finite baryon density, the nucleon masses are reduced due to attractive scalar mean fields, however, the magnetic field continues to induce a substantial enhancement in the effective masses. For instance, at $\rho_B=\rho_0$ and $f_s=0$, the proton mass increases from $751.82$ MeV at $eB=0$ to approximately $792$ MeV at $eB=3\,m_\pi^2$, and further rises to $979.86$ MeV at $eB=5\,m_\pi^2$.
% corresponding to an overall increase of about $30\%$. 
Similarly, the neutron mass increases from $748.66$ MeV at $eB=0$ to $975.03$ MeV at $eB=5\,m_\pi^2$. At lower density ($\rho_B=\rho_0/2$), the trend becomes moderate, where the proton mass increases from about $832$ MeV at $eB=0$ to nearly $1330$ MeV at $eB=5\,m_\pi^2$. These results indicate that magnetic field effects gradually dominate over density effects in the strong magnetic field region. In asymmetric matter ($I_a=0.5$), the proton-neutron mass splitting becomes slightly more significant due to the combined influence of scalar-isovector interactions and magnetic field induced medium modifications. At $\rho_B=\rho_0$ and $f_s=0$, the proton and neutron masses differ by about $3$ MeV at $eB=0$, while the splitting increases slightly with magnetic field and reaches nearly $4-5$ MeV at around $eB=5\,m_\pi^2$, referring to Table~\ref{tab:masses}. Nevertheless, both nucleons exhibit the same characteristic nonlinear increase with magnetic field, indicating that the dominant behavior is still governed by the external magnetic field. The effect of strangeness fraction on the nucleon effective masses remains weaker throughout the considered magnetic field range. For example, at $\rho_B=\rho_0$ and $eB=5\,m_\pi^2$, the proton mass changes only slightly from $979.86$ MeV for $f_s=0$ to $981.61$ MeV for $f_s=0.7$, while the neutron mass changes from $975.03$ MeV to $976.66$ MeV, as shown in Table~\ref{tab:masses}. Thus, the overall magnetic field dependence of the nucleon masses remains nearly unchanged with increasing $f_s$, indicating that magnetic field and density effects dominate over the contribution arising from finite strangeness in the medium. A similar increasing trend with magnetic field is also observed for the hyperons shown in Figs.~\ref{fig:hyperon_1_effective_masses}, ~\ref{Lambda_masses} and ~\ref{fig:hyperon_2_effective_masses}, although the magnitude of the variation depends on their charge and strangeness content, as discussed in the following paragraph.

In Figs.~\ref{fig:hyperon_1_effective_masses}(a)-(f), the magnetic field dependence of the effective masses of the $\Sigma$ hyperons is presented as a function of $eB/m_\pi^2$. Similar to the nucleons, all $\Sigma$ states exhibit a monotonic increase in effective mass with increasing magnetic field strength. The variation remains limited in the region $eB \lesssim 3\,m_\pi^2$, followed by a noticeably stronger behavior at higher magnetic fields, as also reflected in Table~\ref{tab:masses}.
% For example, the vacuum mass of $\Sigma^{+}$ increases from $1193.24$ MeV at $eB=0$ to $1272.73$ MeV at $eB=3\,m_\pi^2$, and further to $1530.39$ MeV at $eB=5\,m_\pi^2$. A similar trend is observed for $\Sigma^{-}$ and $\Sigma^{0}$, indicating that the magnetic-field-induced enhancement remains qualitatively similar for both charged and neutral $\Sigma$ hyperons. Although the charged $\Sigma^\pm$ states couple directly to the magnetic field through Landau quantization, the overall behavior is largely governed by the medium modification of the scalar fields.
At finite density, the masses decrease substantially due to scalar interactions, however, the strong increment at high magnetic field persists. In the case of $\rho_B=\rho_0$ and $f_s=0$, the mass of $\Sigma^{+}$ increases from $1042.14$ MeV at $eB=0$ to $1239.44$ MeV at $eB=5\,m_\pi^2$. The inclusion of isospin asymmetry ($I_a=0.5$) produces only small quantitative changes in the magnetic-field dependence of the $\Sigma$ masses, while the effect of finite strangeness fraction remains negligible throughout the considered magnetic field range. Figs.~\ref{Lambda_masses} (a) and (b) show that the effective mass of the $\Lambda$ hyperon also increases monotonically with magnetic field. The response remains moderate at lower magnetic fields and becomes more evident beyond $eB \sim 3\,m_\pi^2$, similar to the behavior observed for the $\Sigma$ hyperons. As listed in Table~\ref{tab:masses}, the vacuum mass of the $\Lambda$ increases from $1116.24$ MeV at $eB=0$ to $1195.42$ MeV at $eB=3\,m_\pi^2$, and further to $1452.23$ MeV at $eB=5\,m_\pi^2$. Since the $\Lambda$ hyperon is electrically neutral and contains a significant strange-quark component, its magnetic-field dependence arises mainly through medium-induced scalar-field modifications rather than direct Landau quantization effects. At finite density ($\rho_B=\rho_0$), the effective mass decreases to about $961$ MeV at $eB=0$, but still exhibits a substantial increase with magnetic field, reaching nearly $1158$ MeV at $eB=5\,m_\pi^2$. Overall, the magnetic field dependence of the $\Lambda$ remains qualitatively similar in both symmetric and asymmetric matter, with only small quantitative shifts due to isospin asymmetry ($\sim 10$ MeV) and strangeness fraction ($\sim 6$ MeV).

\begin{table}[htbp]
\centering
\renewcommand{\arraystretch}{1.2}

\begin{tabular}{
V
c
V
c|c|c
V
c|c|c
V
c|c|c
V
}

% \begin{tabular}{
% V 
% c V c@{\hspace{0.5cm}}|@{\hspace{0.2cm}}c@{\hspace{0.5cm}}|@{\hspace{0.5cm}}c 
% V 
% c@{\hspace{0.5cm}}|@{\hspace{0.2cm}}c@{\hspace{0.5cm}}|@{\hspace{0.5cm}}c 
% V 
% c@{\hspace{0.5cm}}|@{\hspace{0.2cm}}c@{\hspace{0.5cm}}|@{\hspace{0.5cm}}c 
% V
% }

\noalign{\hrule height 1.5pt}

% \multirow{2}{*}{Masses (MeV)}
\multirow{2}{*}{\shortstack{Masses\\(MeV)}}
& \multicolumn{3}{cV}{$\rho_B = 0$, \quad $I_a = 0$, \quad $f_s = 0$}
& \multicolumn{3}{cV}{$\rho_B = \rho_0$, \quad $I_a = 0.5$, \quad $f_s = 0$}
& \multicolumn{3}{cV}{$\rho_B = \rho_0$, \quad $I_a = 0.5$, \quad $f_s = 0.7$}
\\

\cline{2-10}

& $eB=0$ & $eB=3m_\pi^2$ & $eB=5m_\pi^2$
& $eB=0$ & $eB=3m_\pi^2$ & $eB=5m_\pi^2$
& $eB=0$ & $eB=3m_\pi^2$ & $eB=5m_\pi^2$
\\

\noalign{\hrule height 1.5pt}

$M_p^*$ 
& 939.16 & 1029.19
& 1316.59
& 751.82 & 792.31
 & 979.86
& 751.10 & 797.63 & 981.61
\\
\hline

$M_n^*$ 
& 939.16 & 1028.40 & 1313.85
& 748.66 & 788.81 & 975.03
& 747.62 & 793.90 & 976.66
\\
\hline

$M_\Lambda^*$ 
& 1116.24 & 1195.42 & 1452.23
& 960.96 & 996.09 & 1157.96
& 950.53 & 991.59 & 1151.79
\\
\hline

$M_{\Sigma^0}^*$ 
& 1028.76 & 1271.96 & 1527.68
& 1039.09 & 1073.93 & 1234.70
& 1028.76 & 1069.48 & 1228.56
\\
\hline

$M_{\Sigma^+}^*$ 
& 1193.24 & 1272.73 & 1530.39
& 1042.14 & 1077.32 & 1239.44
& 1032.12 & 1073.09 & 1233.42
\\
\hline

$M_{\Sigma^-}^*$ 
& 1193.23 & 1271.18 & 1524.97
& 1036.04 & 1070.54 & 1229.96
& 1025.41 & 1065.86 & 1223.72
\\
\hline

$M_{\Xi^0}^*$ 
& 1317.65 & 1387.18 & 1616.01
& 1195.48 & 1226.09 & 1365.19
& 1176.00 & 1212.37 & 1351.38
\\
\hline

$M_{\Xi^-}^*$ 
& 1317.65 & 1386.41 & 1613.30
& 1192.45 & 1222.73 & 1360.47
& 1172.66 & 1208.77 & 1346.55
\\

\noalign{\hrule height 1.5pt}

\end{tabular}

\caption{Effective masses of octet baryons are given at baryon densities, $\rho_B$ = 0 and $\rho_0$, isospin asymmetry, $I_a$ = 0 and 0.5, and strangeness fraction $f_s$ = 0 and 0.7.}
\label{tab:masses}
\end{table}
%\end{sidewaystable}
% \textbf{Xi baryons:}

Figs.~\ref{fig:hyperon_2_effective_masses} (a)-(d) present the magnetic-field dependence of the effective masses of the $\Xi^0$ and $\Xi^-$ hyperons. Although these baryons contain two strange quarks, their effective masses still exhibit a substantial increase with increasing magnetic field strength. Similarly to the other members of the octet baryons, the change is moderate initially as a function of $eB/m_\pi^2$ and becomes significantly stronger beyond $eB \sim 3\,m_\pi^2$. The corresponding numerical values are listed in Table~\ref{tab:masses}. The vacuum mass of $\Xi^0$ increases from $1317.65$ MeV at $eB=0$ to $1387.18$ MeV at $eB=3\,m_\pi^2$, and further to $1616.01$ MeV at $eB=5\,m_\pi^2$. Similarly, the $\Xi^-$ mass increases from $1317.65$ MeV to $1613.30$ MeV over the same magnetic field range. At finite density, the effective masses are reduced due to scalar mean-field effects, but the strong magnetic field modification are persistant. At $\rho_B=\rho_0$ and $f_s=0$, the mass of $\Xi^0$ increases from $1195.48$ MeV at $eB=0$ to $1365.19$ MeV at $eB=5\,m_\pi^2$, while the $\Xi^-$ mass increases from $1192.45$ MeV to $1360.47$ MeV. The magnetic field dependence of both cascade hyperons remains nearly identical across all densities and isospin asymmetries, with only small quantitative differences between the neutral and charged states, $\sim3$ MeV for finite density. This indicates that, despite the presence of electric charge in $\Xi^-$, the overall variation of the cascade masses is governed predominantly by medium induced scalar field modifications rather than direct charge dependent effects. The influence of finite strangeness fraction also remains weak. At $\rho_B=\rho_0$ and $eB=5\,m_\pi^2$, the mass of $\Xi^0$ changes only from $1365.19$ MeV for $f_s=0$ to $1351.38$ MeV for $f_s=0.7$, while the $\Xi^-$ mass changes from $1360.47$ MeV to $1346.55$ MeV, as shown in Table~\ref{tab:masses}. Thus, the overall magnetic field dependence of the cascade hyperons remains qualitatively unchanged with increasing strangeness fraction, while the dominant modifications continue to arise from the external magnetic field.
% Figs.~\ref{fig:hyperon_2_effective_masses} (a)-(d), presents the impact of eB on the effective masses of $\Xi^0$ and $\Xi^-$ hyperons. Even though these baryons contain two strange quarks, their masses still show substantial magnetic field induced growth. Both $\Xi^0$ and $\Xi^-$ exhibit nearly identical magnetic field dependence, with their effective masses increasing monotonically and with similar curvature across all densities and isospin asymmetries. No significant qualitative difference between the neutral and charged states is observed, indicating that the magnetic field induced enhancement is dominated by medium effects rather than charge dependent contributions. Table.~\ref{tab:masses} displays the observed values of effective masses at $\rho_B$ = 0 and $\rho_0$, where we can quantitatively compare the effects of $f_s$ at finite $I_a$.

% \textbf{Comparative behavior:}
Across the entire octet, a broadly similar magnetic field response is observed for both charged and neutral baryons, with all states exhibiting a monotonic increase in effective mass as a function of $eB$. Although quantitative differences exist among the baryons, the overall trend does not exhibit a simple ordering solely based on electric charge, indicating that the magnetic field dependence is strongly influenced by medium modifications of the scalar fields in addition to direct charge effects. The variation also shows moderate sensitivity to the strangeness content. Baryons containing none ($p$ and $n$) to single strange quarks ($\Lambda$ and $\Sigma$ hyperons) generally exhibit comparatively larger variations, whereas baryons containing two strange quarks display relatively weaker magnetic field dependence. Nevertheless, the overall qualitative behavior remains similar across the octet. The increase in baryon masses with magnetic field observed in this work agrees with earlier relativistic mean-field studies, where strong magnetic fields modify the system mainly through Landau quantization~\cite{Broderick:2000pe}. Charged baryons show a more noticeable dependence due to their direct coupling to the field. Similar behavior has been reported in CQMF and quark-meson coupling models, where hyperon masses increase with magnetic field without altering mass ordering~\cite{Papazoglou1999,Saito:2005rv}. In NJL-type models, the enhancement of quark masses with magnetic field leads to a corresponding increase in baryon masses through magnetic catalysis~\cite{Klevansky:1992qe,Miransky:2015ava}. The present CQMF results follow the same overall trend, indicating that the increase in octet baryon masses in strong magnetic fields is a common feature across different approaches.

\subsection{Effective Magnetic Moments of Octet Baryons}

\label{Sec_magentic_moment1}

\begin{figure}[htbp]
\centering
% \renewcommand{\thefigure}{4}
% \vspace{-1.5cm}
\begin{subfigure}{0.42\textwidth}
\centering
\captionsetup{justification=centering,labelformat=empty}
\caption{\boldmath$\ \rho_B=\rho_0/2$}
\includegraphics[width=\linewidth]{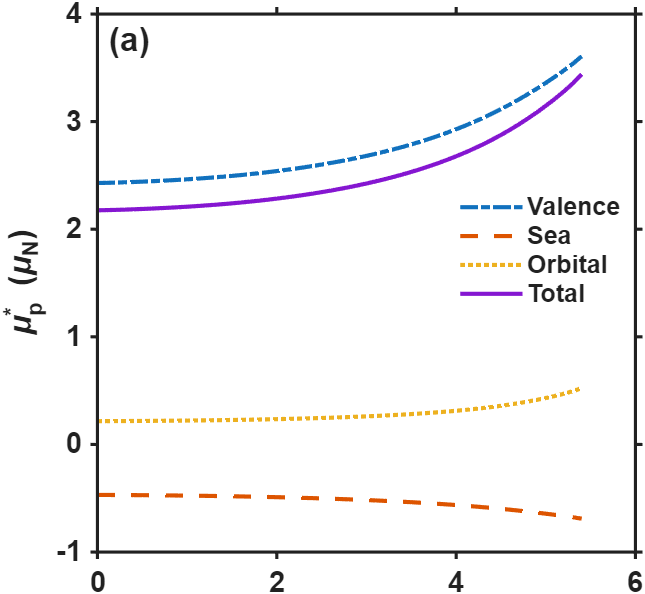}
% \caption{$p,\ \rho=\rho_0/2$}
\end{subfigure}
\hfill
\begin{subfigure}{0.42\textwidth}
\centering
\captionsetup{justification=centering,labelformat=empty}
\caption{\boldmath$\ \rho_B=\rho_0$}
\includegraphics[width=\linewidth]{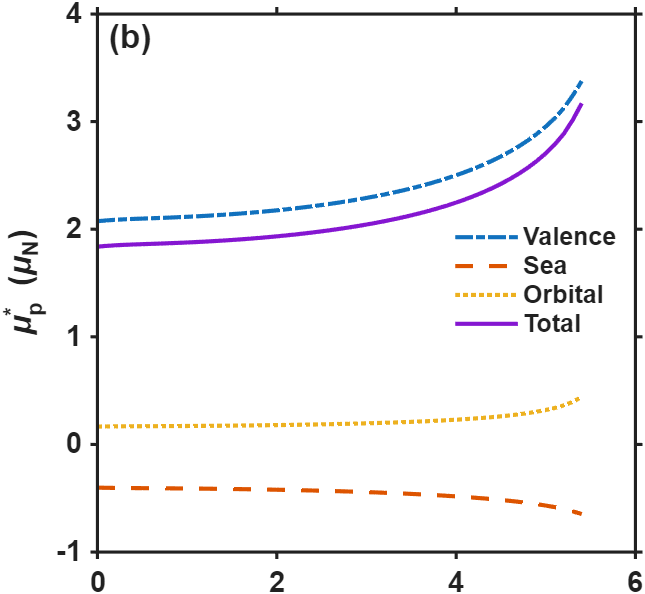}
% \caption{$p,\ \rho=\rho_0$}
\end{subfigure}

\vspace{0.5cm}

\begin{subfigure}{0.42\textwidth}
\centering
\includegraphics[width=\linewidth]{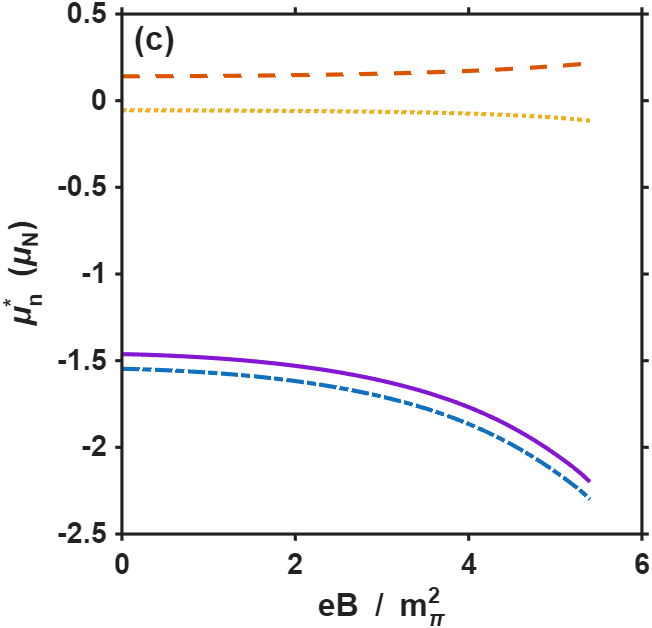}

\end{subfigure}
\hfill
\begin{subfigure}{0.42\textwidth}
\centering
\includegraphics[width=\linewidth]{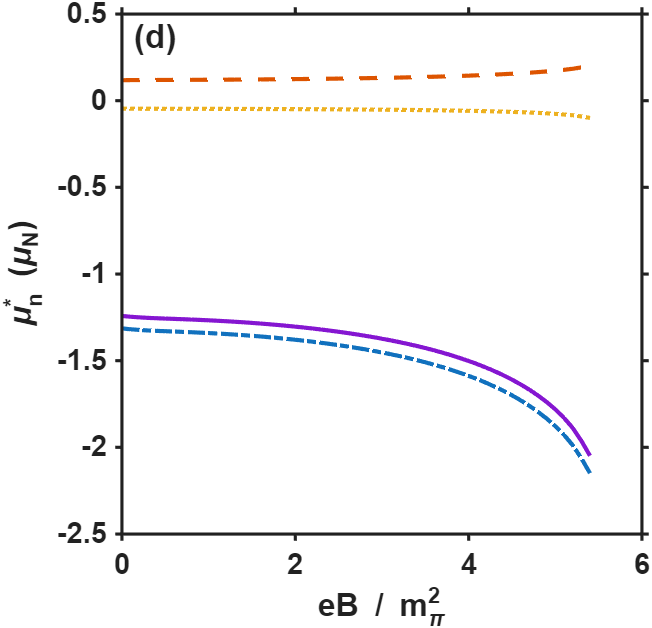}

\end{subfigure}

\caption{Magnetic field dependence of the effective magnetic moments ($\mu_i^*$) of nucleons 
($p$, $n$) at $T=100$ MeV for $I_a=0.5$ and $f_s=0.7$. Subplots (a) and (c) correspond 
to $\rho_B=\rho_0/2$, while subplots (b) and (d) represent results at $\rho_B=\rho_0$. The valence, 
sea, and orbital contributions, along with the total effective magnetic moment, are 
shown as functions of the external magnetic field.}
\label{mm_1}
\end{figure}

\begin{figure}[htbp]
\centering
\vspace{-1cm}
% Lambda

% \vspace{0.5cm}

% Sigma0
\begin{subfigure}{0.42\textwidth}
\centering
\captionsetup{justification=centering,labelformat=empty}
\caption{\boldmath\ $\rho_B=\rho_0/2$}
\includegraphics[width=\linewidth]{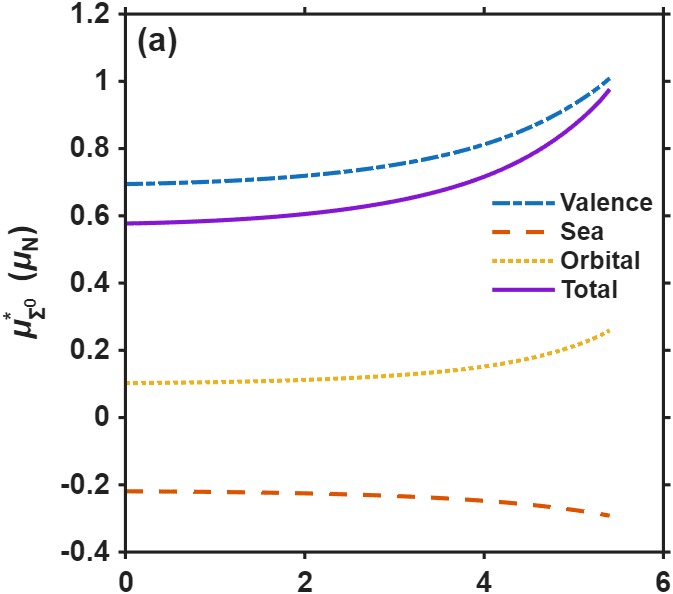}

\end{subfigure}
\hfill
\begin{subfigure}{0.42\textwidth}
\centering
\captionsetup{justification=centering,labelformat=empty}
\caption{\boldmath\ $\rho_B=\rho_0$}
\includegraphics[width=\linewidth]{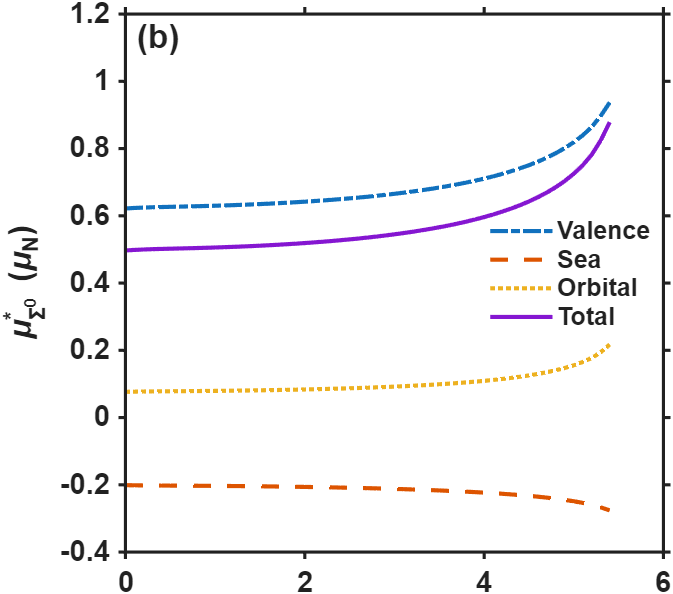}

\end{subfigure}
\vspace{0.5cm}

% Sigma+
\begin{subfigure}{0.42\textwidth}
\centering
\includegraphics[width=\linewidth]{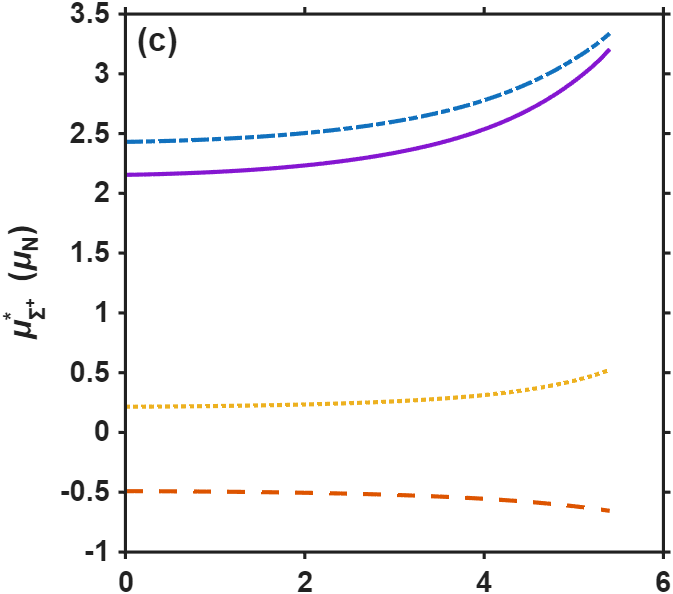}

\end{subfigure}
\hfill
\begin{subfigure}{0.42\textwidth}
\centering
\includegraphics[width=\linewidth]{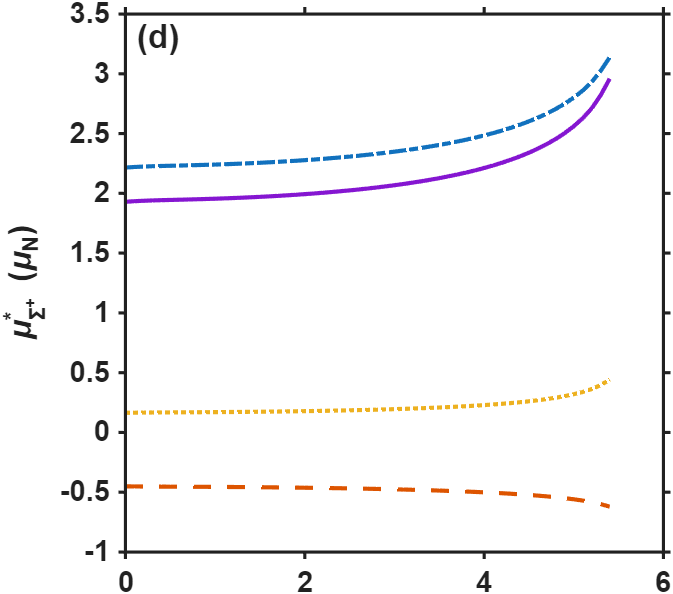}

\end{subfigure}
\vspace{0.5cm}

% Sigma-
\begin{subfigure}{0.42\textwidth}
\centering
% \captionsetup{justification=centering,labelformat=empty}
% \caption{\boldmath\ $\rho_B=\rho_0/2$}
\includegraphics[width=\linewidth]{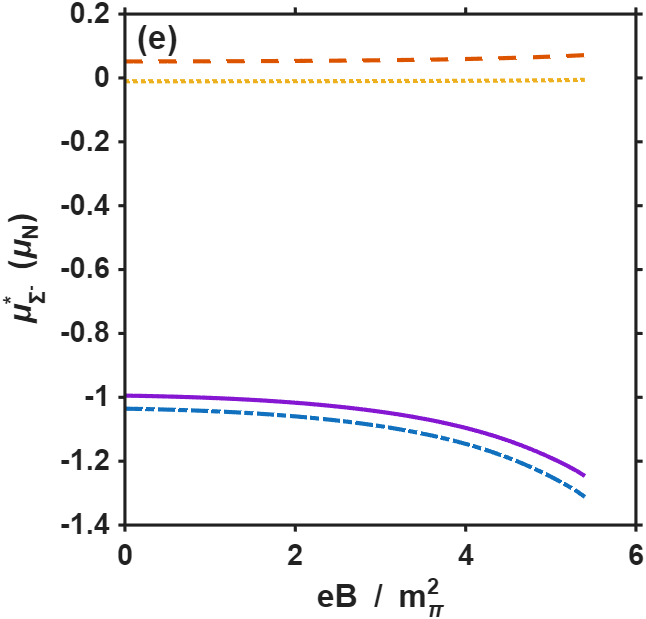}
\end{subfigure}
\hfill
\begin{subfigure}{0.42\textwidth}
\centering
% \captionsetup{justification=centering,labelformat=empty}
% \caption{\boldmath \ $\rho_B=\rho_0$}
\includegraphics[width=\linewidth]{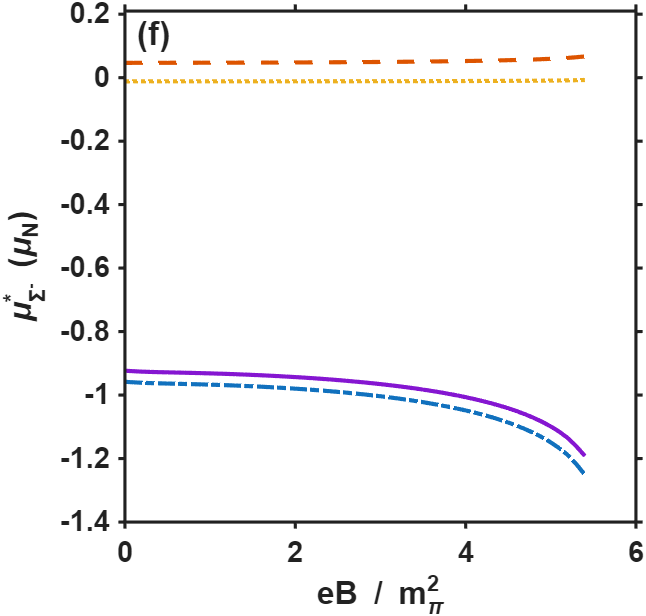}
\end{subfigure}

\caption{Magnetic field dependence of the effective magnetic moments ($\mu_i^*$) of hyperons 
($\Sigma^0$, $\Sigma^+$, $\Sigma^-$) at $T=100$ MeV for $I_a=0.5$ and $f_s=0.7$. 
Subplots (a, c, and e) correspond to $\rho_B=\rho_0/2$, while subplots (b, d, 
and f) represent results at $\rho_B=\rho_0$, representing the contribution of valence, sea, and orbital, along with total effective magnetic moment.
}

\label{fig:hyperon_mu_part1}

\end{figure}

\begin{figure}[htbp]
\centering
% \renewcommand{\thefigure}{5.2}
% \vspace{-1.5cm}

\begin{subfigure}{0.42\textwidth}
\centering
\captionsetup{justification=centering,labelformat=empty}
\caption{\boldmath \ $\rho_B=\rho_0/2$}
\includegraphics[width=\linewidth]{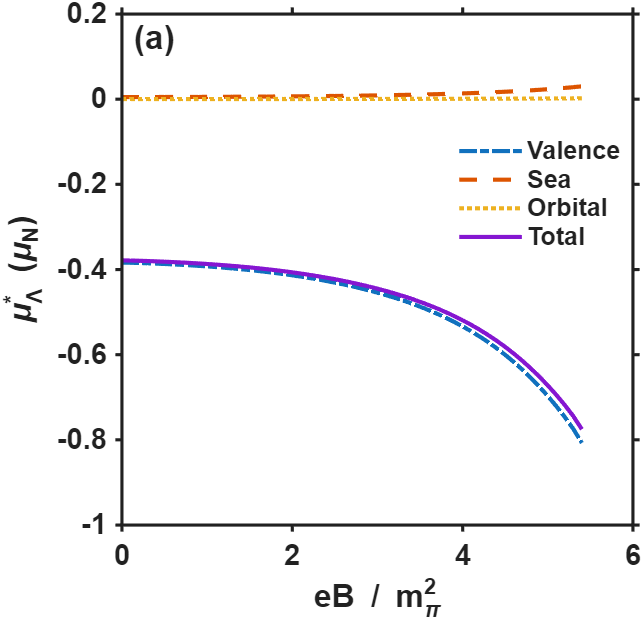}
\end{subfigure}
\hfill
\begin{subfigure}{0.42\textwidth}
\centering
\captionsetup{justification=centering,labelformat=empty}
\caption{\boldmath\ $\rho_B=\rho_0$}
\includegraphics[width=\linewidth]{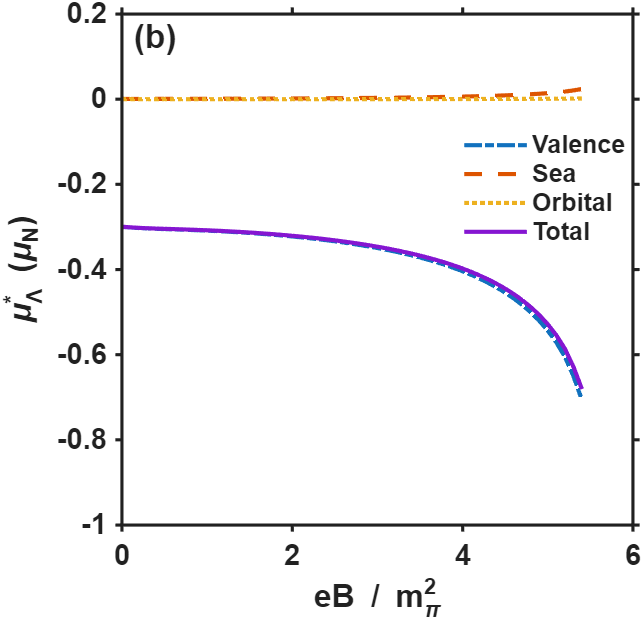}
\end{subfigure}

\caption{Magnetic field dependence of the effective magnetic moments ($\mu_i^*$) of hyperon 
($\Lambda$) at $T=100$ MeV for $I_a=0.5$ and $f_s=0.7$. 
Subplot (a) corresponds to $\rho_B=\rho_0/2$, while subplot (b) represents results at $\rho_B=\rho_0$, showing the contributions of valence, sea, 
and orbital components along with the total effective magnetic moment.
}

\label{Lambda_mm}

\end{figure}

\begin{figure}[htbp]
\centering
% \renewcommand{\thefigure}{5.2}
% \vspace{-1.5cm}

% \vspace{0.5cm}

% Xi0
\begin{subfigure}{0.42\textwidth}
\centering
\captionsetup{justification=centering,labelformat=empty}
\caption{\boldmath \ $\rho_B=\rho_0/2$}
\includegraphics[width=\linewidth]{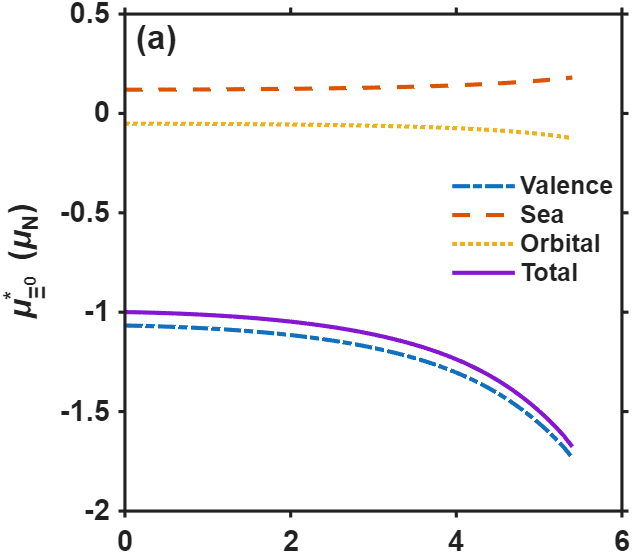}
\end{subfigure}
\hfill
\begin{subfigure}{0.42\textwidth}
\centering
\captionsetup{justification=centering,labelformat=empty}
\caption{\boldmath \ $\rho_B=\rho_0/2$}
\includegraphics[width=\linewidth]{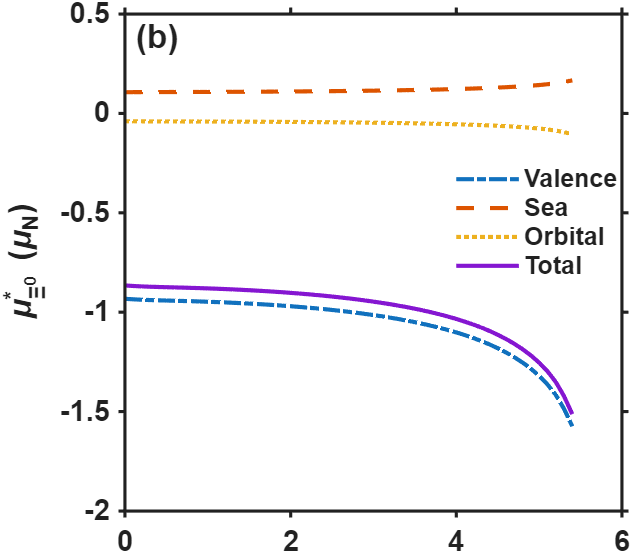}
\end{subfigure}

\vspace{0.5cm}

% Xi-
\begin{subfigure}{0.42\textwidth}
\centering
\includegraphics[width=\linewidth]{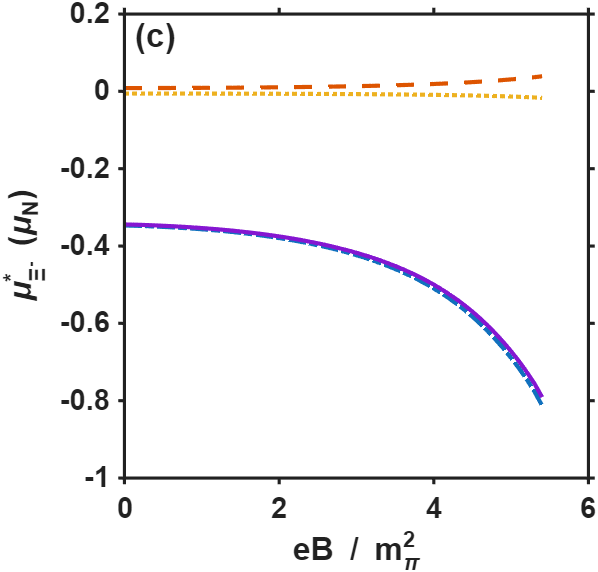}

\end{subfigure}
\hfill
\begin{subfigure}{0.42\textwidth}
\centering
\includegraphics[width=\linewidth]{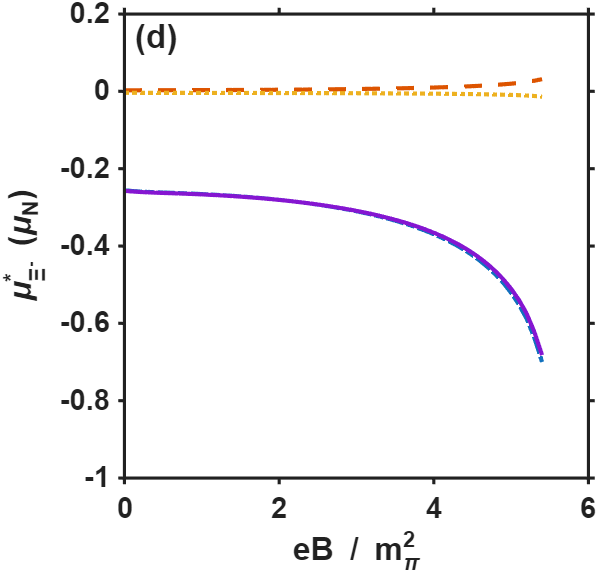}

\end{subfigure}

\caption{Magnetic field dependence of the effective magnetic moments ($\mu_i^*$) of hyperons 
($ \Xi^0$, $\Xi^-$) at $T=100$ MeV for $I_a=0.5$ and $f_s=0.7$. 
Subplots (a) and (c) correspond to $\rho_B=\rho_0/2$, while subplots (b) and (d) represents results at $\rho_B=\rho_0$, showing the contributions of valence, sea, 
and orbital components along with the total effective magnetic moment.
}

\label{fig:hyperon_mu_part2}

\end{figure}

We now investigate the magnetic field dependence of the effective magnetic moments of octet baryons in hot and dense isospin-asymmetric strange matter. The discussion focuses on how the interplay of magnetic field, baryon density, and strangeness content modifies the internal quark structure and spin dynamics of the baryons within the CQMF--$\chi$CQM framework. The calculations are performed at $T = 100$ MeV for fixed isospin asymmetry, $I_a = 0.5$, and strangeness fraction, $f_s = 0.7$. The total magnetic moment is decomposed into valence, sea, and orbital contributions. % \subsection{Effective Magnetic Moments of Octet Baryons:
% Density Dependence} 
Figs.~\ref{mm_1},~\ref{fig:hyperon_mu_part1},~\ref{Lambda_mm} and ~\ref{fig:hyperon_mu_part2} present the effective magnetic moments of octet baryons 
at $I_a=0.5$ and $f_s=0.7$ for baryon densities, $\rho_B=\rho_0/2$ and $\rho_0$. 

Figs.~\ref{mm_1}(a) and (b) show the variation of the proton magnetic moment, $\mu_p^*$, with magnetic field for $\rho_B=\rho_0/2$ and $\rho_0$, respectively. For both the densities, the total magnetic moment increases with $eB$, with larger changes appearing beyond $eB \sim 3\,m_\pi^2$. At $\rho_B=\rho_0/2$, $\mu_p^*$ changes by $0.24$ $\mu_N$ and  $0.92\,\mu_N$  as magnetic field changes from $eB=0$ to $eB=3\,m_\pi^2$, and  $5\,m_\pi^2$, respectively. For $\rho_B=\rho_0$, the corresponding values changes to   $0.18$ $\mu_N$ and  $0.79\,\mu_N$, as seen in Table~\ref{magnetic_moments}. The decomposition of $\mu_p^*$ indicates that the valence quark contribution provides the largest part of the total magnetic moment and follows nearly the same magnetic-field dependence as the total result. The sea contribution remains negative with relatively small variation, whereas the orbital contribution increases gradually with magnetic field, particularly in the high-$eB$ region. Consequently, the overall behavior of the proton magnetic moment is governed mainly by the valence sector. For the neutron, shown in Figs.~\ref{mm_1}(c) and (d), the total magnetic moment remains negative throughout the considered magnetic field region, while its magnitude increases steadily with increasing $eB$. At $\rho_B=\rho_0/2$, magnitude of $\mu_n^*$ increases by $0.14\,\mu_N$ as  $eB$ increases to $3\,m_\pi^2$, and by $0.4\,\mu_N$ over further change of $eB$ from $3\,m_\pi^2$ to $5\,m_\pi^2$. Similarly, for $\rho_B=\rho_0$, the magnetic moment rise is almost similar for the given range. The valence component again dominates the total magnetic moment and becomes increasingly negative with magnetic field, thereby determining the overall trend of $\mu_n^*$. In comparison, the sea contribution remains positive and small in comparison, while the orbital term shows only limited variation across the considered magnetic field interval. Comparing the two density configurations, the absolute values of both proton and neutron magnetic moments become smaller at $\rho_B=\rho_0$ than at $\rho_B=\rho_0/2$. However, the overall dependence on magnetic field remains similar at both densities, indicating that finite density mainly modifies the magnitude of the magnetic moments without significantly changing their magnetic field dependence.

For the $\Sigma$ baryons, shown in Figs.~\ref{fig:hyperon_mu_part1} (a)-(f), the total magnetic moments exhibit a continuous increase in magnitude with magnetic field at both $\rho_B=\rho_0/2$ and $\rho_0$. The valence contribution remains the dominant component for all $\Sigma$ states and largely determines the magnetic field dependence of the total magnetic moments, while the sea contribution stays small and negative throughout the considered $eB$ region. The orbital contribution varies positively with magnetic field for $\Sigma^{0}$ and $\Sigma^{+}$, whereas almost no visible variation is observed in the orbital sector of $\Sigma^{-}$. For the neutral $\Sigma^{0}$ hyperon, shown in Figs.~\ref{fig:hyperon_mu_part1} (a) and (b), the total magnetic moment increases by about $0.06\,\mu_N$ over the region $eB=0$ to $3\,m_\pi^2$ at $\rho_B=\rho_0/2$, followed by a further increase of nearly $0.21\,\mu_N$ up to $eB=5\,m_\pi^2$. Similarly, for $\rho_B=\rho_0$, the magnitude of $\mu_{\Sigma^0}^*$ increases by about $0.05\,\mu_N$ up to $eB=3\,m_\pi^2$, and by an additional $\sim0.16\,\mu_N$ between $eB=3\,m_\pi^2$ and $5\,m_\pi^2$. The stronger variation in the higher magnetic field region mainly follows the behavior of the valence contribution, while the orbital sector provides a smaller positive correction.

The positively charged $\Sigma^{+}$ baryon, shown in Figs.~\ref{fig:hyperon_mu_part1} (c) and (d), exhibits the largest magnetic response among the $\Sigma$ states. At $\rho_B=\rho_0/2$, the total magnetic moment increases by nearly $0.17\,\mu_N$ between $eB=0$ and $3\,m_\pi^2$, and further by $\sim0.56\,\mu_N$ up to $eB=5\,m_\pi^2$. For $\rho_B=\rho_0$, the corresponding increase is about $0.13\,\mu_N$ up to $eB=3\,m_\pi^2$, followed by a further rise of approximately $0.44\,\mu_N$ over the higher magnetic field interval. The large variation observed for $\Sigma^{+}$ originates mainly from the strong magnetic field dependence of the valence sector associated with its charged quark composition. In contrast, the $\Sigma^{-}$ hyperon, shown in Figs.~\ref{fig:hyperon_mu_part1} (e) and (f), retains a negative magnetic moment throughout the considered magnetic field range. At $\rho_B=\rho_0/2$, the magnitude of $\mu_{\Sigma^-}^*$ increases by about $0.11\,\mu_N$ up to $eB=3\,m_\pi^2$, followed by a weaker change of nearly $0.08\,\mu_N$ between $eB=3\,m_\pi^2$ and $5\,m_\pi^2$. Similarly, at $\rho_B=\rho_0$, the magnitude changes by approximately $0.04\,\mu_N$ in the lower magnetic field region and by nearly $0.12\,\mu_N$ at larger $eB$. The valence contribution again governs the total magnetic moment and becomes increasingly negative with magnetic field, whereas the sea and orbital sectors contribute only small corrections. Overall, the magnetic moments of the $\Sigma$ hyperons show stronger variation at $\rho_B=\rho_0/2$ than at $\rho_B=\rho_0$, indicating that finite density suppresses the overall magnetic response without significantly modifying the qualitative magnetic field dependence of the individual valence, sea, and orbital contributions.

The $\Lambda$ hyperon, shown in Figs.~\ref{Lambda_mm} (a) and (b), retains a negative magnetic moment throughout the considered magnetic field region, while its magnitude increases continuously with $eB$. At $\rho_B=\rho_0/2$, the magnitude of $\mu_{\Lambda}^*$ changes by about $0.06\,\mu_N$ between $eB=0$ and $3\,m_\pi^2$, followed by a further variation of nearly $0.21\,\mu_N$ up to $eB=5\,m_\pi^2$. Similarly, at $\rho_B=\rho_0$, the corresponding changes are approximately $0.05\,\mu_N$ and $0.17\,\mu_N$, respectively, indicating that the magnetic field dependence remains qualitatively similar at both densities. The valence contribution dominates the total magnetic moment and becomes increasingly negative with magnetic field, whereas the sea contribution remains positive and limited across the entire $eB$ region. The orbital sector shows only a weak variation with magnetic field and contributes a relatively small correction to the total magnetic moment. Consequently, the overall behavior of $\mu_{\Lambda}^*$ follows mainly the magnetic field dependence of the valence quark sector. For the $\Xi^{0}$ hyperon, shown in Figs.~\ref{fig:hyperon_mu_part2} (a) and (b), the total magnetic moment also remains negative, with its magnitude increasing steadily with magnetic field. At $\rho_B=\rho_0/2$, the magnitude of $\mu_{\Xi^0}^*$ changes by nearly $0.05\,\mu_N$ up to $eB=3\,m_\pi^2$, and by a further $\sim0.42\,\mu_N$ between $eB=3\,m_\pi^2$ and $5\,m_\pi^2$. For $\rho_B=\rho_0$, the corresponding changes are about $0.08\,\mu_N$ and $0.28\,\mu_N$, respectively. The valence contribution remains the dominant component and largely determines the magnetic field dependence of the total magnetic moment, while the sea and orbital contributions exhibit only small variations throughout the considered magnetic field interval. A similar behavior is observed for the $\Xi^{-}$ hyperon in Figs.~\ref{fig:hyperon_mu_part2} (c) and (d), where the magnetic moment remains negative over the entire magnetic field range. At $\rho_B=\rho_0/2$, the magnitude of $\mu_{\Xi^-}^*$ changes by approximately $0.07\,\mu_N$ between $eB=0$ and $3\,m_\pi^2$, followed by an additional variation of nearly $0.24\,\mu_N$ up to $eB=5\,m_\pi^2$. For $\rho_B=\rho_0$, the corresponding variations are about $0.04\,\mu_N$ and $0.19\,\mu_N$, respectively. In both density configurations, the total magnetic moment closely follows the behavior of the valence contribution, whereas the sea and orbital sectors remain negligible. Overall, the $\Xi$ hyperons exhibit a clear sensitivity to the external magnetic field despite their larger strange quark content, while finite density mainly modifies the magnitude of the magnetic moments without substantially altering their overall magnetic field dependence.

\vspace{1cm}
\begin{table}[htbp]
\centering
\renewcommand{\arraystretch}{1.2}

% \begin{tabular}{
% V c V c|c V c|c V c|c V
% }
% \begin{tabular}{
% V c V c@{\hspace{1.5cm}}|@{\hspace{0.5cm}}c 
% V c@{\hspace{1.3cm}}|@{\hspace{0.5cm}}c 
%  V 
% % c@{\hspace{1cm}}|@{\hspace{0.5cm}}c V
% }

\begin{tabular}{
V
c
V
c|c|c
V
c|c|c
V
% c|c|c
% V
}

\noalign{\hrule height 1.5pt}

\multirow{2}{*}{$\mu_B^*$}
& \multicolumn{3}{cV}{$\rho_B = \rho_0/2$, \quad $I_a = 0.5$, \quad $f_s = 0.7$}
& \multicolumn{3}{cV}{$\rho_B = \rho_0$, \quad $I_a = 0.5$, \quad $f_s = 0.7$}
% & \multicolumn{2}{cV}{$\rho_B = \rho_0$, \quad $I_a = 0.5$, \quad $f_s = 0.7$}
\\

\cline{2-7}

& $eB=0$ &  $eB=3m_\pi^2$ & $eB=5m_\pi^2$
& $eB=0$ &  $eB=3m_\pi^2$ & $eB=5m_\pi^2$
% & $eB=0$ & $eB=5m_\pi^2$
\\

\noalign{\hrule height 1.5pt}

$\mu_p^*$ 
& 2.17 & 2.41 & 3.09
& 1.84 & 2.03 & 2.63
% & 751.10 & 981.61
\\
\hline

$\mu_n^*$ 
& -1.46 & -1.60 & -2.00
& -1.24 & -1.36 & -1.74
% & 747.62 & 976.66
\\
\hline

$\mu_\Lambda^*$ 
& -0.38 & -0.44 & -0.65
& -0.29 & -0.34 & -0.51
% & 950.53 & 1151.79
\\
\hline

$\mu_{\Sigma^0}^*$ 
& 0.58 & 0.64
 & 0.85
& 0.49 & 0.54 & 0.70
% & 1028.76 & 1228.56
\\
\hline

$\mu_{\Sigma^+}^*$ 
& 2.15 & 2.32 & 2.88
& 1.93 & 2.06 & 2.50
% & 1032.12 & 1233.42
\\
\hline

$\mu_{\Sigma^-}^*$ 
& -0.99 & -1.10 & -1.18
& -0.92 & -0.96 & -1.08
% & 1025.41 & 1223.72
\\
\hline

$\mu_{\Xi^0}^*$ 
& -0.99 & -1.04 & -1.46
& -0.86 & -0.94 & -1.22
% & 1176.00 & 1351.38
\\
\hline

$\mu_{\Xi^-}^*$ 
& -0.34 & -0.41 & -0.65
& -0.26 & -0.30 & -0.49
% & 1172.66 & 1346.55
\\

\noalign{\hrule height 1.5pt}

\end{tabular}

\caption{The above table showcases the total effective magnetic moment of octet baryons (in units of nuclear magneton $\mu_N$) at baryon densities, $\rho_B = \rho_0/2$ and $\rho_0$, isospin asymmetry, $I_a$ = 0.5 and strangeness fraction $f_s$ = 0.7.}
\label{magnetic_moments}
\end{table}

Across the entire octet, a common trend is observed in which the magnetic moments at $\rho_B=\rho_0/2$ exhibit stronger magnetic field dependence than those at $\rho_B=\rho_0$, as reflected in Table~\ref{magnetic_moments}. This behavior indicates that increasing baryon density tends to moderate the magnetic response of the internal quark structure. The effect is primarily driven by the valence quark sector, which provides the dominant contribution to the total magnetic moments, while the sea and orbital components contribute comparatively smaller corrections. At higher density, modifications induced by the scalar fields and effective quark masses become more pronounced, thereby reducing the sensitivity of quark spin polarization to the external magnetic field. Similar medium effects have been reported in QMC and NJL-based studies, where increasing density leads to substantial modifications of hadron properties through changes in the underlying quark dynamics and chiral condensates \cite{Saito:2005rv,Buballa:2003qv}. Studies based on effective hadronic and bag model frameworks, density and magnetic field induced modifications of hadronic properties have also been discussed in investigations of strongly magnetized hadronic matter and compact star environments \cite{Broderick:2000pe,Chakrabarty1997}. The present CQMF-$\chi$CQM results follow the same qualitative behavior while additionally allowing a decomposition of the magnetic moments into valence, sea, and orbital contributions. The analysis indicates that density effects predominantly influence the valence quark component, whereas the sea-quark contribution remains relatively weak throughout the baryon octet.

\vspace{-0.3cm}
\section{Summary}
\label{Summary}
% \noindent
To summarize the effects of magnetic field, finite baryon density ($\rho_B=0,\rho_0/2,\rho_0$), and strangeness fraction, ($f_s=0$ and 0.7), at finite temperature, $T$ = 100 MeV, together influence the properties of octet baryons within the chiral SU(3) quark mean-field (CQMF) model and the chiral constituent quark ($\chi$CQM) model. The scalar fields determine the in-medium environment and lead to changes in the effective baryon masses. These changes are then reflected in the spin and flavor structure that govern the magnetic moments. It is found that magnetic field effects are more visible at lower densities, whereas higher density tends to moderate the response due to medium effects. The inclusion of strangeness redistributes the contributions between light and strange quark sectors, leading to differences among $\Lambda$, $\Sigma$, and $\Xi$ baryons compared to nucleons ($p$ and $n$). The decomposition of magnetic moments into valence, sea and orbital sector, further indicates that the dominant medium dependence arises from the valence quark contribution, while sea and orbital terms provide smaller corrections. Overall, the combined effect of magnetic field, density, and quark structure leads to a consistent modification of baryon properties, which is relevant for understanding the strongly magnetized dense matter such as that found in compact stars. The present results may also provide useful theoretical guidance for future experimental and astrophysical investigations aimed at exploring the behavior of hadronic matter under extreme magnetic field conditions.

\section*{ACKNOWLEDGMENTS}
H.D. would like to thank  the Science and Engineering Research Board, Anusandhan-National Research Foundation (ANRF), Government of India under the SERB-POWER Fellowship scheme (Ref No. SPF/2023/000116) for financial support.
A.K. sincerely acknowledges Anusandhan-National Research Foundation (ANRF), Government of India for funding of the
research project under the Science and Engineering Research Board-Core Research Grant (SERB-CRG) scheme (File No. CRG/2023/000557).

% \section{Reference}

\end{document}